\documentclass[aps,preprint,groupedaddress]{revtex4}%PRE
\usepackage{amsmath,amssymb,amsthm,mathrsfs,amsfonts,dsfont} 
\usepackage{graphicx}
\usepackage{times}
\usepackage{subfigure}
\usepackage{amsbsy}
\usepackage{bm}
\usepackage{algorithm}
\usepackage{algpseudocode}
\usepackage{algorithmicx}

\usepackage{xcolor}

\newcommand{\blue}[1]{\textcolor{black}{#1}}

\begin{document}

\title{Kinetic energy in random recurrent neural networks}

\author{Li-Ru Zhang$^{1}$}
\author{Haiping Huang$^{1,2}$}
\email{huanghp7@mail.sysu.edu.cn}
\affiliation{$^{1}$PMI Lab, School of Physics,
Sun Yat-sen University, Guangzhou 510275, People's Republic of China}
\affiliation{$^{2}$Guangdong Provincial Key Laboratory of Magnetoelectric Physics and Devices,
Sun Yat-sen University, Guangzhou 510275, People's Republic of China}

\date{\today}

\begin{abstract}
High-dimensional chaotic dynamics can emerge in a large random recurrent neural network when the synaptic gain crosses a threshold. Recent works showed that the kinetic energy of neural activity links the chaotic dynamics and the supporting unstable fixed points (equilibria) in the phase space. Here, we investigate the kinetic-energy-centric properties of random recurrent neural networks by combining dynamical mean-field theory with extensive numerical simulations. 
    We find that the average kinetic energy shifts continuously from zero to a positive value at \blue{the known} critical value of coupling variance (synaptic gain) and exhibits a cubic scaling behavior near the critical point from above. This scaling behavior is supported by numerical simulations and provides a quantitative characterization of how fast the dynamics change during the onset of chaos \blue{as well as how far the chaotic dynamics are away from the unstable fixed points}. The steady-state activity distribution is further calculated by the theory and compared with simulations on finite-size systems from the kinetic-energy optimization perspective as well. The activity distribution is also analyzed in a geometric angle, \blue{revealing that although the original chaotic dynamics and the gradient dynamics of the kinetic energy are arranged in a shell-like structure, they are well separated in the polar direction.} The trajectory length on the chaotic manifold can be derived from the stationary kinetic energy, and the associated stationary behavior is analyzed as well. 
\end{abstract}

\maketitle

\section{Introduction}
High-dimensional recurrent neural networks (RNNs) with random synaptic couplings provide a minimal yet powerful theoretical framework for understanding nonequilibrium dynamics, emergent computation, and collective fluctuations in complex systems.
 Increasing synaptic gain induces a continuous transition from a stable fixed point to high-dimensional chaos~\cite{Chaos-1988}. This intriguing collective behavior makes the random RNN a paradigmatic model in statistical physics, neuroscience, and machine learning. A standard tool for analyzing this chaos transition is dynamical mean-field theory (DMFT)~\cite{PRE-2018,Roy-2019,CMP-2024,Zou-2024,Carles-2024,Exact-2025}, which offers an exact description of the non-equilibrium steady dynamics in the thermodynamic limit.

RNNs with random couplings exhibit a continuous dynamics phase transition slightly above which the dynamics are extremely slow but sensitive to external perbubations~\cite{Chaos-1988,Zou-2024,Qiu-2025}.
The dynamics phase transition thus shows self-organized criticality via tuning only the variance of synaptic coupling despite different forms of synaptic connectivity, which offers computational advantage for reservoir computing~\cite{EoC-1990,EoC-2004,Maass-2009,Takasu-2025}. The model also provides a theoretical basis for better understanding population computation in real brain circuits~\cite{Abbott-2009,Kadmon-2015,Mastrogiuseppe-2018,Vyas-2020,Huang-2024,Ostojic-2024}. However, understanding the nature of chaotic dynamics in RNNs
remains incomplete. A hot current debate is about whether the unstable fixed points after the chaos transition are related to the chaotic dynamics. Recent works argued that the fixed point ensemble provides transiently attractive landmarks that
guide the chaotic flow~\cite{Helias-2025,HWZ-2025}. Moreover, fixed points and dynamics are confined to separate shells in the phase space~\cite{Helias-2025,Wang-2024,Yang-2025}. Nevertheless, other works challenged this viewpoint and argued that the high-dimensional chaos can not be inferred from these equilibria, showing theoretical evidence in a mixture model of conservative and non-conservative forces~\cite{Urbani-2025}.

Inspired by searching for fixed points (equilibria) in RNNs~\cite{NN-2013}, we formulated the fixed point distribution as an optimization problem of minimizing kinetic energy of the dynamics in a recent work~\cite{Qiu-2025}. \blue{It was recently shown that unstable fixed points are reachable by considering an Onsager feedback term in the original non-gradient dynamics~\cite{HWZ-2025}. Furthermore, the kinetic energy landscape can be used to drive a generative model named neural Langevin machine~\cite{Yu-2025}. In addition, the kinetic energy serves as a natural metrics measuring how far the chaotic dynamics is away from the unstable fixed points. Therefore, it is interesting and strongly motivated to study how the kinetic energy varies with the synaptic gain and the associated impact on the dynamics on the chaotic attractor.} 

By applying dynamical mean-field theory~\cite{PRE-2018,Urbani-2025}, we find that at the chaos transition point, the stationary (in the long time limit) kinetic energy becomes positive in a continuous way, and we determine the scaling exponent for the critical behavior close to the transition point. A cubic growth is identified. Interestingly, this behavior is shared with the linear dimensionality growth~\cite{Clark-2023} as well as the topological complexity growth~\cite{Helias-2025}, differently from the quadratic growth of the maximum Lyapunov exponent during the onset of chaos~\cite{Chaos-1988,PRR-2023}. The theory is corroborated by numerical simulations. Moreover, the single-time probability distribution of the chaotic activity is also derived and calculated by the DMFT, supported by simulations in a finite-size system and compared with the gradient dynamics under the kinetic energy as well. The chaotic steady state is mapped to stationary dynamics of a Langevin-type gradient system with an effective temperature fixed by kinetic-energy matching. Both systems share the same distribution of synaptic currents, yet their activities are located at different shells with a spherical rotation (revealed in finite system simulations). Whether this holds in infinite-sized systems remains to be explored in the future. Finally, we show that the arc length of chaotic trajectories grows linearly with a rate determined by the stationary kinetic energy. We expect this work will improve our theoretical understanding of neural computation at the onset of chaos and far beyond the edge of chaos.

\section{Model setting and DMFT}
This section reviews the standard dynamical mean-field theory for random recurrent neural networks, following previous works~\cite{Chaos-1988,PRE-2018,Helias-2020,Huang-2022}, since the kinetic energy relies on the DMFT framework. The part particularly related to the derivation of kinetic energy and its asymptotic behavior is our contribution in the current work and will be mentioned in the next section. 

\begin{figure*}
    \centering
    \includegraphics[width=0.9\textwidth]{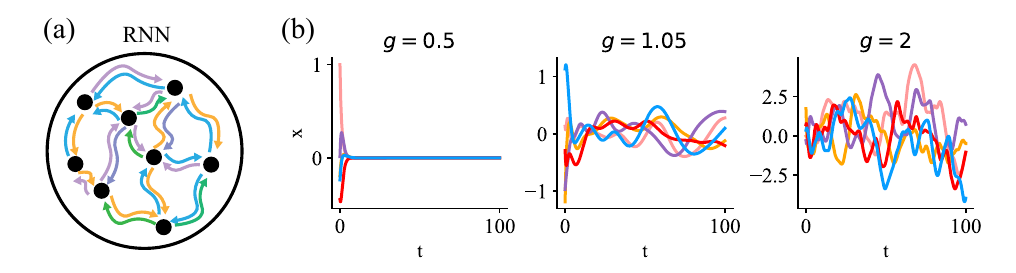}
    \caption{ Sketch of network connectivity and dynamics. (a) Schematic illustration of the RNN connectivity structure. (b) Sample trajectories of five neurons for coupling strengths $g = 0.5,1.05,2$ in a neural pool of $N=5\,000$.}
        \label{fig1}
\end{figure*}

We consider a recurrent neural network composed of  $N$ interacting units whose dynamics evolve according to the following equation:
\begin{equation}
    \partial_t x_i(t)= -x_i(t)+ \sum_{j}  J_{ij}\phi(x_j),
    \label{neuron_dynamics}
\end{equation}
where $x_i(t)$ denotes the synaptic current  [transformed to activity through $\phi(\cdot)$] of neuron $i$ at time $t$, $\phi(x)$ is a nonlinear activation function [in this work we use $\text{tanh}(\cdot)$], and $J_{ij}$ are synaptic weights drawn independently from a Gaussian distribution with zero mean and variance  $\frac{g^2}{N}$.
Here, $g$ is also referred to as the synaptic gain, which controls the overall level of recurrent synaptic strength in the network.
When $g$ crosses a threshold of $g_c = 1$, a dynamics phase transition is triggered from a fixed point of null activity to chaos~\cite{Chaos-1988}. Slightly above the critical coupling strength $g_c$, the system enters an unstable regime, where the dynamics become highly sensitive to small perturbations and spontaneously evolve into a dynamics regime of chaotic fluctuations~(Fig.~\ref{fig1}).

In the limit of a large network size ($N\rightarrow \infty$), we apply the framework of dynamical mean-field theory to reduce the high-dimensional system to an effective single-unit description. Under DMFT, the dynamics of a representative neuron are governed by the following stochastic differential equation:
\begin{equation}
    \partial_t x(t) = -x(t) + \eta (t),
    \label{DMFT_Dynamics}
\end{equation} 
where $\eta(t)$ is a Gaussian random process of zero mean that encapsulates the cumulative effect of the recurrent inputs from the rest of the network. The second-order statistics are thus determined self-consistently by the network's own dynamics:
\begin{equation}
    \langle \eta(t) \eta(t^\prime)\rangle_\eta = g^2 C_{tt^\prime} ,
\end{equation}
where $C_{tt'}$ is the autocorrelation function, defined as $C_{tt^\prime} = \langle \phi(t)\phi(t^\prime) \rangle_\eta$, where $\phi(t)\equiv\phi(x(t))$. 

Assuming that the system reaches a stationary regime, we consider the time-translationally invariant (TTI) limit where both $t,t^\prime \rightarrow \infty$, and only the time lag ($\tau=t-t'$) matters. In the TTI limit, we have the following motion equation:
\begin{equation}
    (1-\partial_\tau^2)\Delta_\tau = g^2 C_\tau,
    \label{partial_tau^2_Delta}
\end{equation}
where $C_\tau\equiv\langle\phi(t)\phi(t+\tau)\rangle$, and $\Delta_\tau=\langle x(t)x(t+\tau)\rangle$. \blue{The derivation of Eq.~\eqref{partial_tau^2_Delta} is briefly summarized in Appendix~\ref{DMFT_Derivation}, although more technical details can be found in recent textbooks~\cite{Helias-2020,Huang-2022}.}
In the following analysis, we further distinguish between the equal-time and time-lagged ($\tau$-lagged) correlations,  $\Delta_0$ and $\Delta_\tau=\Delta$, respectively.

Since $x(t)$ is a zero-mean Gaussian process whose covariance matrix is a function of $\Delta$ and $\Delta_0$, the nonlinear correlation function $C_{tt^\prime}$ can be written as~\cite{Huang-2022}:
\begin{equation}
    C(\Delta; \Delta_0) = \int Dz  \left[ \int Dx \phi (x \sqrt{\Delta_0 - |\Delta|} +z\sqrt{|\Delta|}) \right]^2,
    \label{C_D_D0}
\end{equation}
where $Dx$ and $Dz$ denote the integration measure over standard Gaussian random variables.
To provide an intuitive picture of the system's evolution, we introduce an effective potential function  $V(\Delta ;\Delta_0)$, such that:
\begin{equation}\label{meq}
    \partial_\tau^2\Delta = -\frac{\partial V(\Delta ; \Delta_0)}{\partial \Delta   
    }.
\end{equation}
This leads to the explicit form:
\begin{equation}
    \begin{aligned}
     V& \left(\Delta ;\Delta_0\right)=\\
    &-\frac{\Delta^2}{2} +{g^2} \int D z\left[\int D x \Phi\left( x \sqrt{\Delta_0-|\Delta|}+ z \sqrt{|\Delta|}\right)\right]^2 .
    \label{V_D_D0}
\end{aligned}
\end{equation}
where  $\Phi(x )= \int_0^x \mathrm{d} y \phi(y) $. It can be verified that Eq.~\eqref{partial_tau^2_Delta} and Eq.~\eqref{meq} are the same by by applying the Price's theorem~\cite{Price-1958}. A detailed derivation of the DMFT equations and the associated potential function can be found in Appendix~\ref{DMFT_Derivation}. In the following section, we use this theoretical framework to investigate the kinetic energy properties and probability distributions of neural activity in RNNs.

\section{Kinetic energy analysis}
In this section, we present \blue{how to derive and analyze the average kinetic energy in random RNNs, which relies on the motion equation in Eq.~\eqref{partial_tau^2_Delta} (see the previous section)}. We also derive the scaling law of kinetic energy in the critical regime where $g \rightarrow 1^+$. Furthermore, we examine the stationary distribution of RNN activity obtained from DMFT in comparison with that obtained through gradient dynamics on kinetic energy keeping an equivalent temperature, and their geometric locations are also compared,  and finally we analyze the arc length of neural trajectories, which depends on the steady state kinetic energy and provides insights into the temporal evolution of the network dynamics in the steady state regime.

\subsection{Average kinetic energy }
\blue{The kinetic energy is defined as the $\ell_2$ norm of the state velocity, namely $\frac{1}{2}\mathbf{v}^2$, where $\mathbf{v}$ is the velocity [see Eq.~\eqref{neuron_dynamics}].
The average kinetic energy quantifies how far away the chaotic dynamics are from those unstable fixed points of zero speed in typical observations.
Therefore, the kinetic energy serves as a quantitative measure of the dynamics landscape. In the following analysis, we neglect the prefactor $1/2$ for convenience. Hence, the average kinetic energy is the expected squared velocity.}
Starting from Eq.~\eqref{DMFT_Dynamics}, we first obtain the following speed correlation:
\begin{equation}
    \begin{aligned}
        \Gamma_{tt^\prime}
        &\equiv\left\langle\partial_t x(t) \partial_{t^{\prime}}  x\left(t^{\prime}\right)\right\rangle \\
        & =\int \frac{\mathrm{d} \omega}{2 \pi} \int \frac{\mathrm{d} \omega^{\prime}}{2 \pi} \omega \omega^{\prime}\left\langle\hat{x}(\omega) \hat{x}^*\left(\omega^{\prime}\right)\right\rangle e^{-i \omega t} e^{i \omega^{\prime} t^{\prime}} \\
        & \overset{\text{TTI}}{=}\int \frac{\mathrm{d} \omega}{2 \pi} \int \frac{\mathrm{d} \omega^{\prime}}{2\pi}  \omega^2 2\pi\delta(\omega - \omega^\prime )\hat{\Delta } (\omega)  e^{-i \omega t} e^{i \omega^{\prime} t^{\prime}} \\
        & =\int \frac{\mathrm{d} \omega}{2 \pi} \omega^2 \hat{\Delta}(\omega)e^{-i \omega\tau}  \\
        & =-\partial_\tau^2 \Delta_\tau ,
       \label{KE_TTI}
    \end{aligned}
\end{equation}
where the stationary limit is taken, $\tau=t-t'$, and the Fourier transformation is applied (technical details can be found in Appendix~\ref{DMFT_Derivation}). This speed correlation was first explored in a different context~\cite{Urbani-2025}. Here we give an alternative derivation of $\Gamma_{tt^\prime} = -\partial_\tau^2 \Delta_\tau$ by Fourier transformation. \blue{The Fourier transformation trick is widely used in analyzing stochastic dynamics~\cite{Risken-1996,Van-2007,Helias-2020,Huang-2022}.} 

 According to Eq.~\eqref{partial_tau^2_Delta}, which relates $\partial_\tau^2 \Delta_\tau$ to $C_\tau$, \blue{we derive the average kinetic energy below. First, the TTI limit of $\Gamma_{tt^\prime}$, i.e., $\Gamma(\Delta; \Delta_0)$ is given by
\begin{equation}\label{dV}
    \Gamma(\Delta;\Delta_0) =\frac{\partial V(\Delta;\Delta_0)}{\partial \Delta }=g^2C(\Delta;\Delta_0) - \Delta.
\end{equation}
Second, taking the limit $\tau \rightarrow 0$, we obtain the average kinetic energy at equal times ($\tau=0$)}:
\begin{equation} \label{Gamma_0}
    \Gamma(\Delta=\Delta_0;\Delta_0)\equiv\Gamma(\Delta_0) = g^2C(\Delta_0) - \Delta_0
    =g^2 \int Dz \left[  \phi (z\sqrt{|\Delta_0|}) \right]^2 - \Delta_0.
\end{equation}
In the long-time limit, the velocity at distant time separation becomes uncorrelated:
\begin{equation}
    \lim\limits_{\tau \rightarrow \infty} \Gamma(\tau) = 0.
\end{equation}

To compute the average kinetic energy, one should first determine the value of $\Delta_0$. To accomplish this, we should notice that
thanks to the motion equation [Eq.~\eqref{meq}], the energy is conserved at the initial and final state, i.e., $V(\Delta_0;\Delta_0)=V(0;\Delta_0)$.  \blue{Therefore, we solve the following energy-conservation equation: $\mathcal{F}(\Delta; \Delta_0)=V(\Delta;\Delta_0)-V(0;\Delta_0)$ evaluated at $\Delta = \Delta_0$ vanishes, where the explicit expression of $\mathcal{F}(\Delta;\Delta_0)$ is given below.}
\begin{equation}
\begin{aligned}
    &\mathcal{F}(\Delta;\Delta_0) \\
    &= -\frac{\Delta^2}{2} 
    +{g^2} \int D z\left[\int D x \Phi\left( x \sqrt{\Delta_0-|\Delta|}+ z \sqrt{|\Delta|}\right)\right]^2  \\
    &- g^2\left[ \int Dz  \Phi\left(\sqrt{\Delta_0}{z}\right)\right]^2 ,
\end{aligned}
\end{equation}
which allows us to compute the DMFT prediction for the average kinetic energy. We use a Python-based numerical solver to find the solution of $\mathcal{F}(\Delta_0; \Delta_0) = 0$ (energy conservation), and then substitute the resulting $\Delta_0$ back into Eq.~\eqref{Gamma_0}. \blue{In the following sections, we denote $\Gamma(\Delta_0)$ as $\Gamma_0$.}

\begin{figure}[t]
    \centering
    \includegraphics[width=0.95\linewidth]{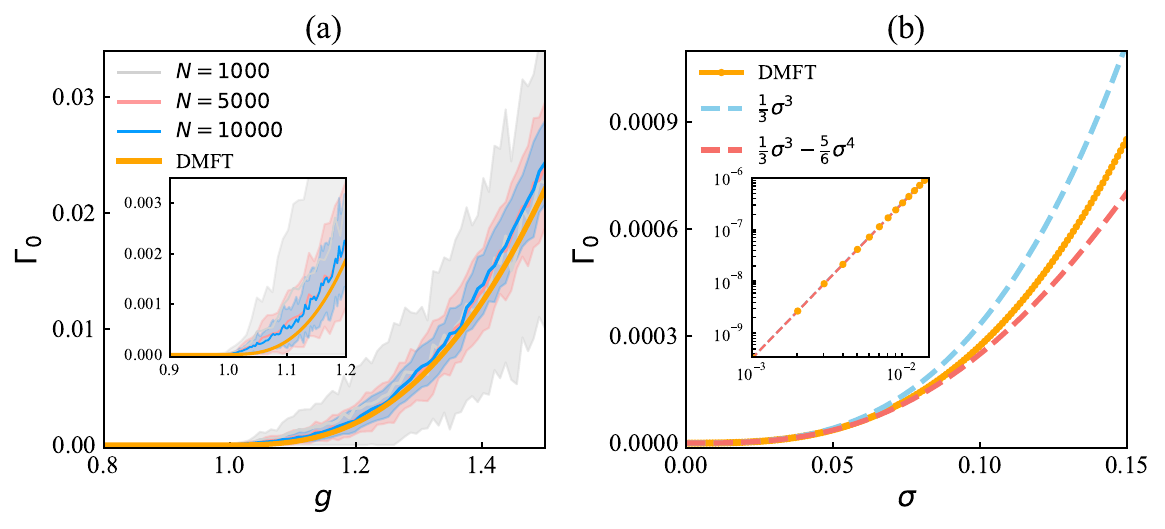}
    \caption{
   \blue{ The average kinetic energy of RNNs in the stationary limit and its asymptotic behavior around the critical point.
    (a) Stationary kinetic energy $\Gamma_0$ as a function of the synaptic gain $g$.
    Colored curves show direct simulations for network sizes $N=1000,5000,10000$, and shaded regions denote sample-to-sample fluctuations over independent coupling-realizations.
    The orange curve is the DMFT result, which is obtained by numerically solving $\mathcal{F}(\Delta_0; \Delta_0) = 0$ for $\Delta_0$, and substituting the solution into Eq.~(\ref{Gamma_0}).
    The inset magnifies the near-critical regime.
    (b) Critical behavior for $g=1+\sigma$.
    The DMFT result is compared with the leading asymptotic law $\Gamma_0=\sigma^3/3$ and the fourth-order expansion $\Gamma_0=\sigma^3/3-5\sigma^4/6$.
    The inset shows the log-log comparison close to the transition.}
    }
    \label{fig2}
    \end{figure}

Figure~\ref{fig2} (a) compares the average kinetic energy obtained from direct simulation of Eq.~\eqref{neuron_dynamics} with the DMFT prediction from Eq.~\eqref{Gamma_0}. The plot  shows results for a wide range of coupling strengths $0.8 \leq g \leq 1.5$. We observe excellent agreement between the simulation results and the DMFT theory. Moreover, the fluctuations in the simulated data decrease as the network size increases, as expected.

\subsection{Critical behavior of kinetic energy at $g\rightarrow 1^+$}
In the critical limit $g \rightarrow 1^+$, the equal-time correlation $\Delta_0$ tends to zero. Because $\Delta$ is bounded by $\Delta_0$~\cite{Huang-2022}, 
$|\Delta| \ll 1$ holds for all $\tau$ as $\sigma = g - 1 \rightarrow 0^+$. In this regime, we expand both the potential function $V(\Delta; \Delta_0)$ and the energy-conservation function $\mathcal{F}(\Delta; \Delta_0)$ in powers of $\Delta$ and $\Delta_0$, and retain terms up to the leading nontrivial (fourth) order:
\blue{
\begin{equation}
    \begin{aligned}
    V(\Delta;\Delta_0) &\thicksim g^2\left( \frac{1}{4}\Delta_0^2 -\frac{1}{4}\Delta_0^3 + \frac{19}{48}\Delta_0^4-\frac{1}{6}\Delta_0^5 \right) \\
    &+ \left[ -1+g^2(1-2\Delta_0 +5\Delta_0^2-\frac{46}{3}\Delta_0^3) \right]\frac{\Delta^2}{2}\\
    &+g^2 \frac{\Delta^4}{6}-\frac{4}{3}g^2\Delta_0\Delta^4 ,
    \label{V_expansion}
    \end{aligned}
\end{equation}
}
which separates into a constant term (independent of $\Delta$) and a $\Delta$-dependent part. As discussed previously, $\mathcal{F}(\Delta; \Delta_0)$ is defined as the difference between the potential evaluated at an arbitrary $\Delta$ and its value at the fixed point $\Delta = 0$. Therefore, by substituting $\Delta = \Delta_0$ into Eq.~\eqref{V_expansion}, we obtain the expansion of $\mathcal{F}(\Delta; \Delta_0)$:
\blue{\begin{equation}
    \mathcal{F}(\Delta_0;\Delta_0) \thicksim \left[ -1+g^2(1-2\Delta_0 +5\Delta_0^2-\frac{46}{3}\Delta_0^3) \right]\frac{\Delta_0^2}{2}+g^2\frac{\Delta_0^4}{6}-\frac{4}{3}g^2\Delta_0^5.
    \label{F_0}
\end{equation}
}
 According to Eq.~\eqref{F_0}, we solve $ \mathcal{F}(\Delta_0;\Delta_0)=0$ for $\Delta_0$ near the critical point $g = 1 + \sigma$, where $\sigma \rightarrow 0^+$, obtaining the following expansion:
 \blue{
\begin{equation}
  \Delta_0 \thicksim \sigma +\frac{7}{6}\sigma^2 -\frac{7}{9}\sigma^3 + \mathcal{O}(\sigma^4).
    \label{Delta0_expan}
\end{equation}
}
\blue{Technical details of the above expansions are given in Appendix~\ref{scaling_derivation}. An expansion up to the second order leads to $\Delta_0 \thicksim \sigma +\frac{7}{6}\sigma^2$, as can be seen from Eq.~\eqref{Delta0_expan}. Here, we obtain $\Delta_0$ following the method presented in Ref.~\cite{PRE-2018}, with a slightly different form of dynamics [see Eq.~\eqref{neuron_dynamics}]. We remark that the RNN dynamics studied in Ref.~\cite{PRE-2018} different from ours yields the same critical point with ours, but different asymptotic behaviors of both kinetic energy and $\Delta_0$ when higher order corrections are considered. We give more details in 
Appendix~\ref{scaling_derivation} (note that details have not been shown in Ref.~\cite{PRE-2018}).
Equation~\eqref{Delta0_expan} will be used below to determine the scaling behavior of the kinetic energy when $g\to g_c^+$.}

By expanding the right-hand-side of Eq.~\eqref{dV} or directly taking the derivative of Eq.~\eqref{V_expansion} with respect to $\Delta$, we obtain:
\blue{
\begin{equation}\label{gamma0asyt}
    \Gamma_0  \thicksim(- \frac{\sigma^2}{3}+\frac{11}{9}\sigma^3)\Delta+(1-6\sigma)\frac{2\Delta^3}{3}.
\end{equation}
}
Using Eq.~\eqref{Delta0_expan}, we derive the asymptotic behavior of the kinetic energy near the critical point $g \rightarrow 1^+$:
\blue{
\begin{equation}\label{gamma0exp}
    \Gamma_0  \thicksim \frac{1}{3}\sigma^3 -\frac{5}{6}\sigma^4 + \mathcal{O}(\sigma^5).
\end{equation}
}
The detailed derivation of these results is provided in Appendix~\ref{scaling_derivation}. 
From the above derivations in this section, we obtain an analytical result for the kinetic energy, which scales with $\sigma$ as a cubic growth law when $g\to g_c$ from above.
As shown in Fig.~\ref{fig2} (b), the theoretical prediction of the kinetic energy in the critical regime is in excellent agreement with the numerical solution of the DMFT equations. Our theoretical result [Eq.~(\ref{gamma0exp})] begins to deviate from the DMFT prediction as the gain factor $g$ increases, because the expansion is only valid in the vicinity of the critical point [Fig.~\ref{fig2} (b)]. \blue{We also show in the figure that as a higher-order correction is taken, the accuracy of the asymptotic result improves.}
This confirms that the scaling law of the kinetic energy near the criticality follows roughly a cubic behavior, with the theoretical scaling exponent equal to $3$. 

The nature of \blue{chaotic activity} can also be understood by other physical quantities.
The maximal Lyapunov exponent quantifies the exponential separation rate of nearby trajectories and has been shown to grow quadratically with $\sigma$~\cite{Chaos-1988,PRR-2023}.
By contrast, the linear dimensionality of chaotic attractors bears a cubic growth slightly above the transition~\cite{Clark-2023}, and the topological complexity of unstable fixed points has also a cubic growth~\cite{Helias-2025}. Interestingly, the kinetic energy also displays a cubic growth just above the critical point. They all may have a common root in the dynamics and fixed-point landscape. It is promising in future works to elucidate this hidden relationship.

\subsection{Single-time probability distribution and geometric structure of stationary activity}
Since $x(t)$ is a Gaussian process with a fixed initial condition (chosen to be zero here), its single-time probability distribution at any time $t$ is fully characterized by a Gaussian distribution with mean $\langle x(t) \rangle$ and variance $\Delta_{tt}$. This distribution can be expressed as~\cite{Exact-2025}:
\begin{equation}
    P(x(t))=\frac{1}{\sqrt{2\pi \Delta_{tt}}}\exp \left[- \frac{(x-\langle x(t) \rangle)^2 }{2\Delta_{tt}} \right].
\end{equation}
The mean is zero, as indicated by the following formal solution:
\begin{equation}
    x(t) = \int_{0}^t \mathrm{d}s e^{-(t-s)}\eta(s).
\end{equation}
The variance is determined self-consistently by solving the single-unit DMFT equations [Eq.~\eqref{DMFT_Dynamics}, see also Appendix~\ref{Algorithm}].

\begin{figure}[htbp]
    \centering
    \includegraphics[width=0.95\textwidth]{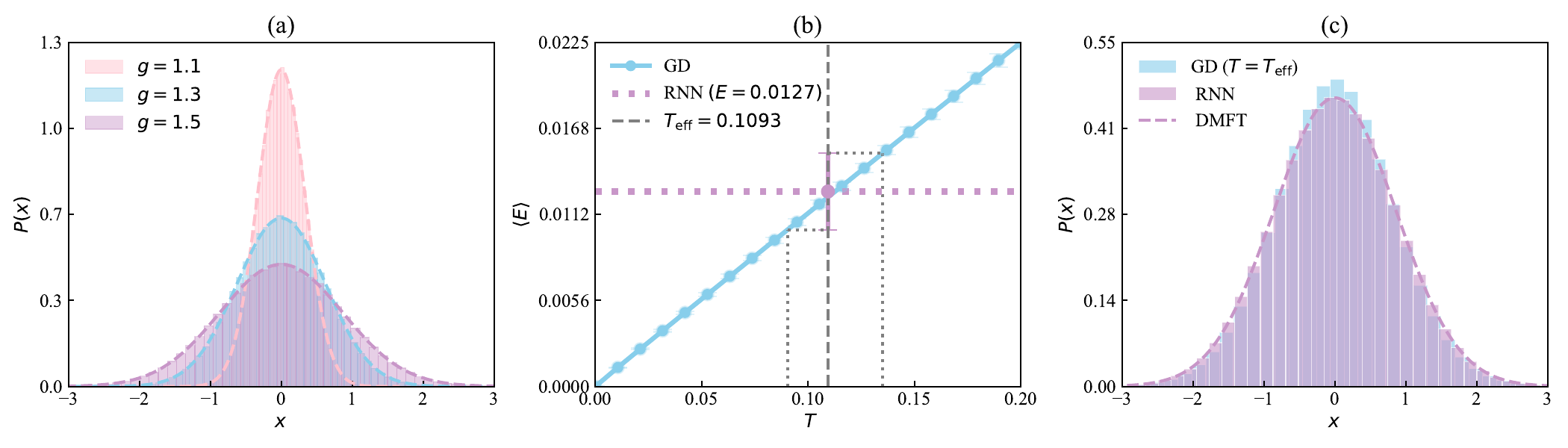}
    \caption{Probability distribution of stationary activity.
        (a) Distribution of $x(t)$ at steady states for different coupling strengths $g=1.1,1.3,1.5$. Histograms represent numerical simulation results from RNN simulations  ($N=5000$), while dashed lines correspond to the Gaussian distributions obtained by DMFT.
        (b) Mean kinetic energy as a function of temperature ($g=1.5,N=5000$). The blue solid curve represents the steady-state average kinetic energy of the gradient descent (GD) dynamics in Eq.~\eqref{GD_dynamics}. 
        The purple dashed curve shows the corresponding steady-state average for the RNN dynamics in Eq.~\eqref{neuron_dynamics} with the same interaction matrix. The vertical dashed line indicates the effective temperature (mean matched with the RNN counterpart).
        The mean kinetic energy and the associated error bars (only shown for the matched point) are computed from $200$ statistically independent samples taken in the stationary regime.    
             (c) Comparison between sampled steady state activities of the two dynamics (the GD one at the effective temperature). $g=1.5$, and $N=5000$. 
        }
    \label{Fig4}
\end{figure} 

To verify this theoretical prediction, we simulate RNN dynamics ($N=5000$) and collect samples of $x(t)$ at the steady state of the system as an empirical distribution. As shown in Fig.~\ref{Fig4}, the probability density obtained from the DMFT framework (represented by the dashed lines) agrees remarkably well with the histograms derived from large-scale numerical simulations of the RNN dynamics. Furthermore, as the coupling strength $g$ increases, the width of the distribution gets broadened, indicating a strong effect of chaotic fluctuations in the network activity. This observation is also in accord with the kinetic energy growth analyzed before.

\begin{figure}[htbp]
    \centering
    \includegraphics[width=0.85\textwidth]{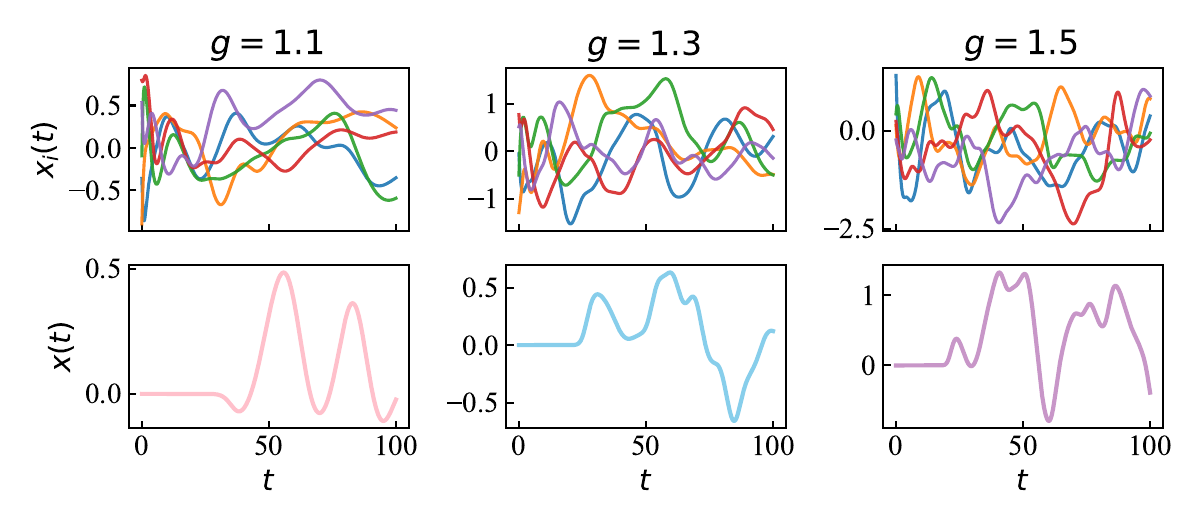}
    \caption{Dynamical trajectories of the RNN (network size $5000$) and the corresponding trajectories from the DMFT equations. Parameters from left to right: $g=1.1,1.3,1.5$.}
    \label{fig5}
\end{figure}

In Fig.~\ref{fig5}, the dynamical trajectories of the RNN and the corresponding trajectories from the DMFT equations are shown for different values of $g$. As $g$ increases, the fluctuations in the trajectories become more pronounced. The algorithmic steps for solving the DMFT equation [Eq.~\eqref{DMFT_Dynamics}] can be found in Appendix~\ref{Algorithm}. This result provides an intuitive understanding about why the distribution of stationary activity gets broadened in Fig.~\ref{Fig4} and about why the arc length growth rate increases with $g$ in the next section.

Next, we explore how the stationary activity is geometrically distributed. We consider a kinetic energy optimization, which will be compared with the original non-gradient RNN dynamics \blue{[see Eq.~\eqref{neuron_dynamics}]}. Introduced in an earlier work~\cite{Qiu-2025}, the kinetic-energy optimization dynamics read
\begin{equation}
    \dot{\mathbf{x}} = -\nabla_{\mathbf{x}} E(\mathbf{x}) + \sqrt{2T}\,\boldsymbol{\epsilon}(t),
    \label{GD_dynamics}
\end{equation}
with the Gaussian white noise $\boldsymbol{\epsilon}(t)$ and the kinetic energy
\begin{equation}
  E(\mathbf{x}) = \frac{1}{2} \sum_{i=1}^N \left( -x_i + \sum_{j=1}^N J_{ij} \phi(x_j) \right)^2.
\end{equation}
By construction, this gradient system admits an equilibrium Boltzmann distribution~\cite{Risken-1996}
\begin{equation}
  P(\mathbf{x}) = \frac{1}{Z} \exp\left( -\beta E(\mathbf{x}) \right),
\end{equation}
where $\beta = \frac{1}{T}$ is related to the noise stength in Eq.~\eqref{GD_dynamics}.

We compute the average energy $\langle E\rangle$ as a function of temperature $T$ for the gradient system and compare the result with the mean kinetic energy of the original RNN in the stationary state ($E = \frac{1}{2}\Gamma_0$), as shown in Fig.~\ref{Fig4} (b).
This comparison defines an effective temperature $T_{\mathrm{eff}}$ via a bisection procedure, such that the mean kinetic energy of the two dynamics coincides.
Importantly, $T_{\mathrm{eff}}$ is not introduced as a free parameter, but determined by the intrinsic fluctuation strength generated by chaotic dynamics.

When the gradient dynamics is simulated at $T_{\mathrm{eff}}$, the resulting stationary single-neuron activity distribution matches that of the original RNN, as demonstrated in Fig.~\ref{Fig4} (c).
This result shows that, at the level of single-time statistics, the nonequilibrium stationary state of the chaotic RNN can be mapped onto an equilibrium-like ensemble characterized by an effective temperature.
From this perspective, the chaotic fluctuations play a role analogous to thermal noise in a corresponding gradient system.

\begin{figure}[htbp]
    \centering
    \includegraphics[width=0.98\textwidth]{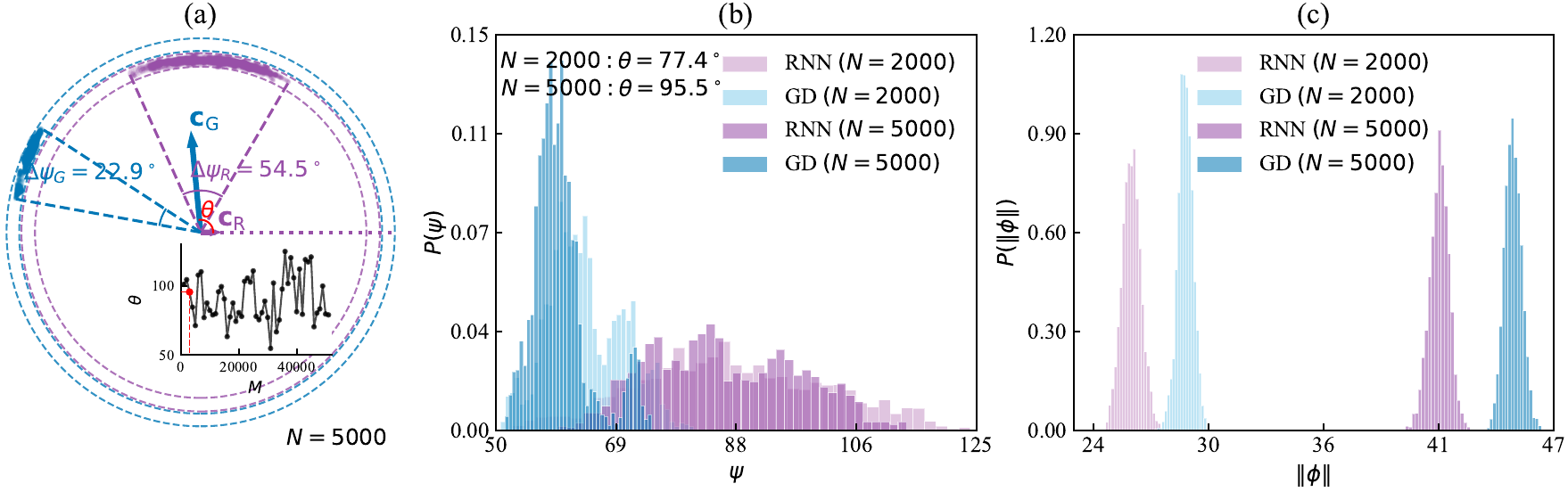}
    \caption{ 
        Geometric structure of stationary activity for RNN and gradient dynamics at $g = 1.5$.
        (a) Two-dimensional geometric representation of stationary activity samples for system size $N=5000$.
        Each point indicates a single high-dimensional state projected onto a polar representation with radius $\left\|\phi \right\|$ and angle $\psi$ with the ensemble average center $c_{\rm G}$ or $c_{\rm R}$.
        Purple (RNN) and blue (GD) clouds correspond to the original RNN dynamics and the associated gradient dynamics, respectively.
        Solid arrows indicate the mean vectors $\mathbf{c}_{R}$ and $\mathbf{c}_{G}$, while dashed lines mark the realm of the distributions. 
        \textbf{Inset}: Angle $\theta$ between the center vectors of the RNN and gradient dynamics as a function of the number of stationary samples $M$.
        The samples are obtained by uniformly sampling different time points along a single stationary trajectory for each dynamics, using the same realization of the coupling matrix and identical initial conditions. The red point indicates that $M=3\,000$ is used for the current cloud representation.
        (b) Probability desity $P(\psi)$ of the angular coordinate $\psi$ for system sizes $N=2000$ and $N =5000$. 
        For each system size, the gradient dynamics are simulated at the corresponding effective temperature, i.e.,
        $T_{\mathrm{eff}}=0.1140 ~ (N=2000)$ and $T_{\mathrm{eff}}=0.1093~(N=5000)$.
        The two dynamics exhibit distinct angular dispersion and a finite misalignment angle $\theta$ between their centroid directions.
        (c) Probability density of the radial norm $\left\|\phi \right\|$. All samples are taken from a single long trajectory in the stationary regime in order to probe the structure of the concrete attractor.
        }
    \label{Fig6}
\end{figure}

An immediate remaining question is whether both dynamics occupy the same part of the phase space. To address this question, we perform the following analysis. 
We consider the original RNN dynamics (denoted by R) and the associated gradient dynamics (denoted by G), using the same realization of the coupling matrix $\mathbf{J}$ and identical initial conditions. As assumed, the dynamics will cover the entire manifold given an infinite simulation time.
For the gradient dynamics, the temperature is chosen to be equal to the effective temperature extracted from the RNN dynamics [see Fig.~\ref{Fig4} (b)].

Starting from the common initial state, each system is evolved in time until entering the stationary regime, as verified by the changes of macroscopic observables such as kinetic energy. In the stationary regime, we record system states every fixed time interval, thereby reducing temporal correlations between successive samples. In total, we collect $M=3\,000$ stationary samples from each dynamics, yielding two ensembles $\{\phi(\mathbf{x})_{R}^\nu\}$ and $\{\phi(\mathbf{x})_{G}^\nu\}$, $\nu=1,2,\cdots,M$. From each ensemble, we compute the centroid vector:
\begin{equation}
  \mathbf{c}_{R} = \big\langle \phi(\mathbf{x})_{R} \big\rangle_M, \quad
\mathbf{c}_{G} = \big\langle \phi(\mathbf{x})_{G} \big\rangle_M,
\end{equation}  
where $\left\langle \cdot \right\rangle_M $ denotes an average over the $M$ stationary samples. For each sample, we then define the polar angle with respect to its corresponding centroid vector as follows~\cite{Yang-2025}
\begin{equation}
  \psi_R^\nu = \arccos\left( \frac{\phi(\mathbf{x})_R^\nu \cdot \mathbf{c}_R}{\|\phi(\mathbf{x})_R^\nu\| \|\mathbf{c}_R\|} \right), \quad
\psi_G^\nu = \arccos\left( \frac{\phi(\mathbf{x})_G^\nu \cdot \mathbf{c}_G}{\|\phi(\mathbf{x})_G^\nu\| \|\mathbf{c}_G\|} \right);
\end{equation}
and also calculate the angle between the two centroid vectors:
\begin{equation}
  \theta = \arccos\left( \frac{\mathbf{c}_R \cdot \mathbf{c}_G}{\|\mathbf{c}_R\| \|\mathbf{c}_G\|} \right).
\end{equation}

As illustrated in Fig.~\ref{Fig6}, the stationary states of the two dynamics form dispersed clouds in the phase space, organized around their respective centroids.
The angular dispersion observed in Fig.~\ref{Fig6} (b) corresponds to the  $\Delta{\psi}$ illustrated in Fig.~\ref{Fig6} (a), while the activity $\ell_2$ norm distribution $\|\phi\|$ is relatively narrow and shown in Fig.~\ref{Fig6} (c) corresponding to the difference between the radii of the outer and inner circular envelopes in Fig.~\ref{Fig6} (a). Figure~\ref{Fig6} (b) shows that the angular distributions 
differ markedly between the two systems, with a rotation of polar angle [$\theta$ in Fig.~\ref{Fig6} (a)]. 
The inset of Fig.~\ref{Fig6} (a) shows that even when $M$ is increased up to $2\times 10^4$, the angle $\theta$ does not converge to a fixed value but instead exhibits persistent fluctuations over a broad range.
Although both dynamics are arranged in a shell-like structure, they are well separated in the polar direction. Whether this geometric property holds in the thermodynamic limit deserves further study in future theoretical works. \blue{In the zero temperature limit, the gradient dynamics will approach the intrinsic unstable fixed-points of the original RNN~\cite{Qiu-2025}, but these fixed points repel the original chaotic dynamics~\cite{Helias-2025}}. A recent work studied an inhibition-dominated neural system and found that the dynamical cap always lies inside
the region containing the fixed points~\cite{Yang-2025}.

\subsection{ Arc length parametrization of RNN activity}
As illustrated in Fig.~\ref{fig5}, representative trajectories of the RNN and the corresponding DMFT dynamics are shown for different values of the coupling strength $g$. The network activity $\bm{x}(t)\in\mathbb{R}^N$ evolves on the chaotic attractor with fluctuations increasingly pronounced as $g$ is increased. This is related to the growth of the kinetic energy, which can be linked to the arc length of trajectories studied in this section.  As time flows, the activity travels along a curve on the manifold embedded in an ambient space of $\mathbb{R}^N$. The arc length is thus the cumulative distance traveled from $t_0$ to $t$, depending on an integral of speed (related to the kinetic energy).  We can compare the results with predictions from DMFT. The arc length $s(t)$ from an initial time $t_0$ to time $t$ is thus defined as follows:
\begin{equation}
    s(t)=\int_{t_0}^t \sqrt{\frac{1}{N}\sum_{i=1}^N\left(\frac{d x_i(\tau)}{d \tau}\right)^2} d \tau,
\end{equation}
where the speed is taken as a population average.

\begin{figure}[htbp]
    \centering
    \includegraphics[width=0.8\textwidth]{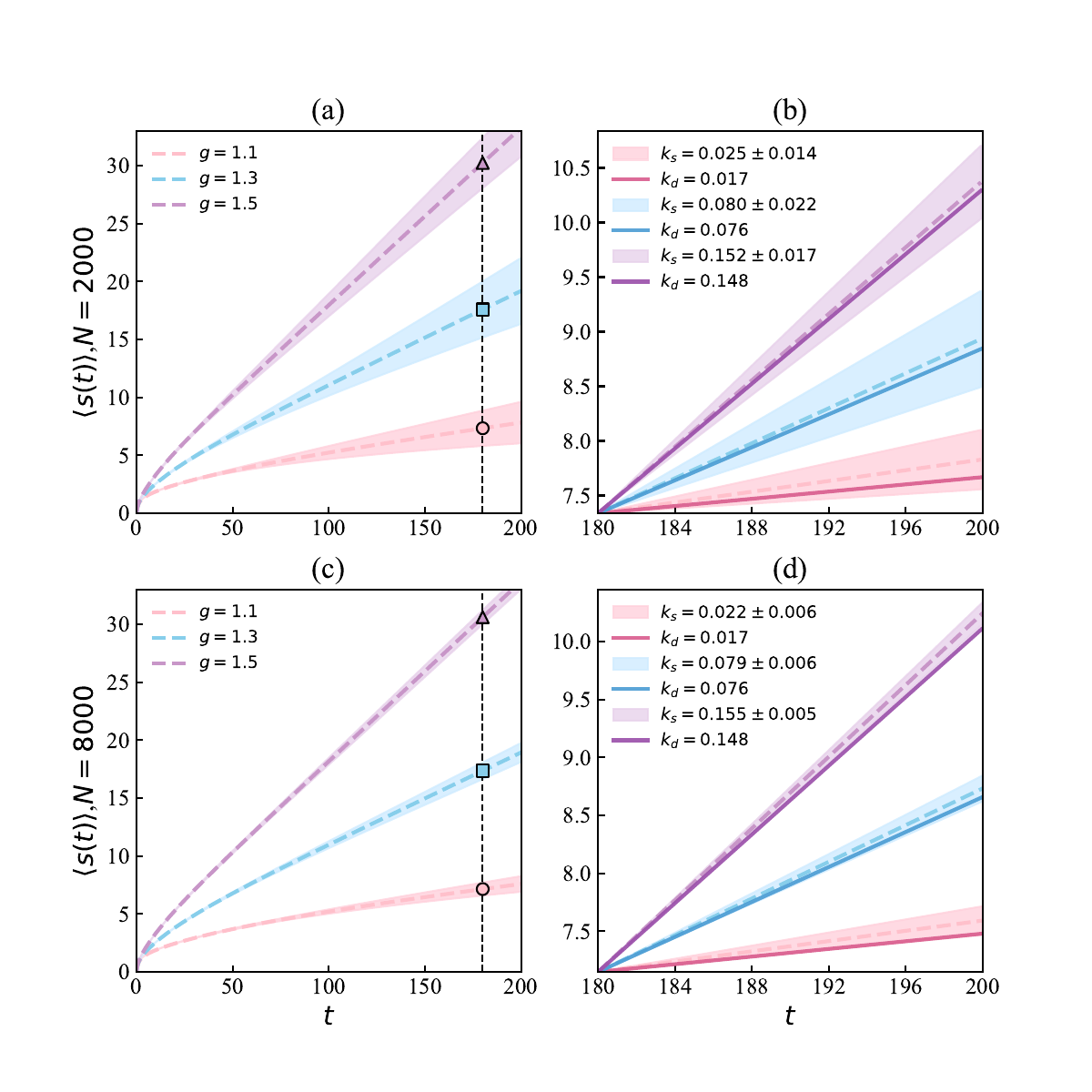}
    \caption{ 
        Arc length of RNN trajectories and comparison with DMFT results. For simulations, two network sizes are considered. In simulations, $\mathrm{d}t=0.1$ is the time unit. (a, c) Arc length computed for RNNs with coupling strengths $g=1.1,1.3,1.5$ for two different network sizes, averaged over $100$ realizations. (b, d) The slope $k_s$ of arc length obtained by a linear fitting over the last $200$ time steps ($\Delta t=200\times dt$) is compared with the DMFT-predicted slope $k_d=\sqrt{\Gamma_0}$ from Eq.~\eqref{Gamma_0}. Shaded regions represent standard deviations across $100$ trajectories. Focusing on the slopes, we fix the intercept $c$ in the linear fit $s(t)=kt+c$ to the height of the corresponding marker in (a) or (c).}
    \label{Fig7}
\end{figure} 

Figure~\ref{Fig7} (a,c) shows the arc length obtained from direct RNN simulations for different coupling strengths $g=1.2,1.3,1.5$ and for network sizes $N=2000$ and $N=8000$.
Each curve (dashed line) is averaged over $100$ independent realizations.
In the chaotic regime, $s(t)$ grows monotonically with time, and its growth rate increases with $g$, reflecting the growth of dynamical fluctuations.
After a transient stage, the system reaches a statistically stationary state in which the arc length exhibits an approximately linear growth $s(t) \simeq k_s\,t$ (subtracted by a baseline), indicating a constant average speed in state space.

In Fig.~\ref{Fig7} (b,d), we perform a linear fit to the arc length trajectory over the final interval 
$t \in [180,200]$. The resulting slope $k_s$ is extracted for different values of $g$. According to the theoretical framework presented earlier, in the steady state, this slope is expected to match the square root of the system's kinetic energy $\sqrt{\Gamma_0}$, predicted by DMFT. We thus compare the fitted slope $k_s$ of the simulated arc length with the DMFT prediction $k_d  = \sqrt{\Gamma_0}$, where  $\Gamma_0$ is computed via Eq.~\eqref{Gamma_0} using the self-consistent solution of $\mathcal{F}(\Delta_0; \Delta_0) = 0$ . 
As shown in Fig.~\ref{Fig7} (b,d), the shaded regions represent the standard deviation over $100$ finite-sized RNN simulations, confirming that the arc length is directly related to the kinetic energy of the system and the estimation error is suppressed by increasing network size. The agreement between the fitted slopes and DMFT predictions further validates this connection. The discrepancies between simulation and theory arise in the case of $g$ close to $g_c$, probably due to numerical precision in simulating the RNN dynamics.

\begin{figure}[htbp]
    \centering
    \includegraphics[width=0.8\textwidth]{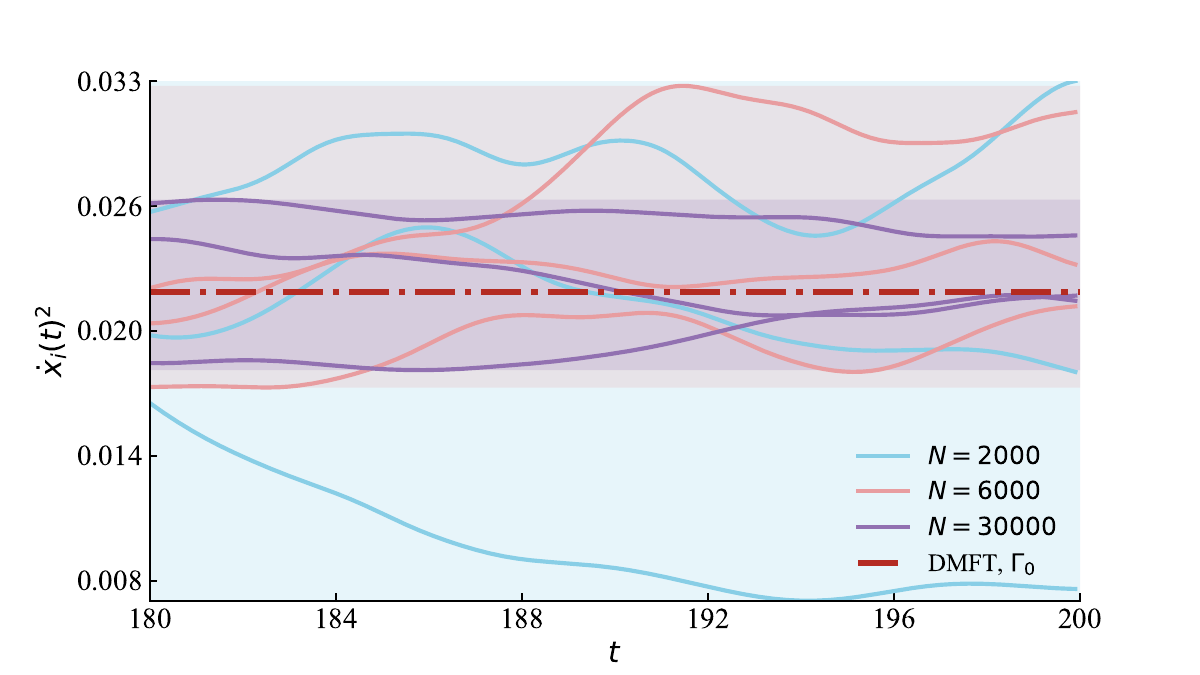}
    \caption{\blue{Fluctuations of the single-neuron squared velocity $\dot{x}^2_i(t)$ in the stationary regime for different network sizes $N =2\,000,6\,000,30\,000$. The synaptic gain is set to $g=1.5$. Shaded regions indicate the range between the maximum and minimum values obtained from the selected (single neuron) trajectories for different network sizes.The dash-dotted line indicates the DMFT prediction $\Gamma_0$.}
    }
    \label{Fig8}
\end{figure}

To further elucidate the origin of slight deviations between simulations and DMFT predictions, we study the temporal fluctuations of the \blue{single neuron squared-velocity $\dot{x}_i^2(t)$} for different network sizes at fixed coupling $g=1.5$, as shown in Fig.~\ref{Fig8}.
While the time-averaged value converges to the DMFT prediction, finite-size networks exhibit strong temporal fluctuations that decrease with increasing $N$.
These fluctuations translate into variations of the instantaneous speed and hence into fluctuations in the measured arc-length growth rate [see simulation results in Fig.~\ref{Fig7} (b,d)].
We expect that in the thermodynamic limit, such fluctuations vanish, and the linear relation between arc length and the square root of kinetic energy becomes exact.

\section{Conclusion}

In this work, we study the nature of high-dimensional neural dynamics in terms of kinetic energy within the framework of dynamical mean-field theory.
We characterize the chaotic regime through the stationary mean kinetic energy, which quantifies the typical speed of network trajectories in phase space in the stationary regime.
By solving the DMFT equations in the stationary limit, we show that the average kinetic energy emerges continuously at the well-known transition to chaos and exhibits a cubic growth close to the critical coupling.
This scaling behavior is consistent with numerical simulations and provides a coherent description of the onset of chaos, being of the same critical exponents as linear dimensionality of chaotic attractors and topological complexity~\cite{Clark-2023,Helias-2025}.

By comparing the stationary activity statistics of the original random recurrent network with those generated by a gradient-descent dynamics with the same kinetic energy, we clarify through finite-size network simulations that the activity statistics are approximately the same, but in terms of geometric organization, they are different, highlighting the necessity to derive theoretical arguments in the thermodynamic limit.  

The arc length \blue{measuring how long} the neural trajectory flows along the chaotic manifold can also be linked to the stationary kinetic energy predicted by DMFT. In particular, the arc length grows linearly in time in the stationary regime, with a rate determined by the stationary kinetic energy. 
This establishes a direct link between microscopic dynamics and the emergent properties of chaotic attractors in high-dimensional neural systems.
From a computational perspective, the arc-length-based characterization of chaotic manifolds may provide useful insights toward understanding neural state representations in cognitive circuits~\cite{Fiete-2019,Fusi-2023,Miller-2025}.

\blue{Our current work is limited to random recurrent neural networks without any learning. It is thus promising to study the kinetic energy landscape of recurrent neural network with ongoing synaptic plasticity, such as those studied in recent works~\cite{Clark-2024,Du-2024,Jstat-2023} and fixed-points statistics used to learn real data~\cite{Yu-2025}.}

\begin{acknowledgments}
 We are grateful to two referees for their constructive comments. This research was supported by the National Natural Science Foundation of China for Grant numbers 12475045, and Guangdong Provincial Key Laboratory of Magnetoelectric Physics and Devices (No. 2022B1212010008), and Guangdong Basic and Applied Basic Research Foundation (Grant No. 2023B1515040023).
\end{acknowledgments}

\section*{Code availability}
Codes to reproduce all results are deposited in our Github~\cite{ZLR-2025}.

\appendix

\section{Details of DMFT Derivation}
\label{DMFT_Derivation}
This part provides a self-contained derivation of known DMFT results for completeness and pedagogical clarity.

\blue{First, the effective noise $\eta(t)$ in Eq.~\eqref{DMFT_Dynamics} is not white, but temporally correlated.
 This mean-field process can be derived by applying the cavity method in a self-consistent way~\cite{Zou-2024}. In the original network, the input current to neuron $i$ is given by $\sum_j J_{ij} \phi(x_j(t))$. Within the DMFT framework, the input is modeled as a Gaussian process whose statistics are self-consistently 
determined by one-dimensional effective dynamics. In essence, the Gaussian description emerges from a cavity-type argument in the large-network
 limit or from a path-integral formalism (see technical details in a lecture note~\cite{Zou-2024}). }

A formal solution to Eq.~(\ref{DMFT_Dynamics}) reads:
\begin{equation}
    x(t) = \int_{0}^t \mathrm{d}s e^{-(t-s)}\eta(s).
\end{equation}
Since $x(t)$ is a zero-mean Gaussian process determined by its covariance matrix $\boldsymbol{\Delta}$, the nonlinear correlation function $C_{tt'}$ can be written as:
\begin{equation}
        C_{tt^\prime} = \int\int \frac{\mathrm{d}\mathbf{x}}{2\pi \sqrt{\det  \boldsymbol{\Delta}}}\exp\left[- \frac{1}{2}\mathbf{x}^T  \boldsymbol{\Delta}^{-1} \mathbf{x} \right] \phi(x(t))\phi(x(t^\prime)) , 
\end{equation}
where we define $ \mathbf{x}^T =  \left( x(t),x(t^\prime) \right)$, and the covariance matrix $\boldsymbol{\Delta} \in \mathbb{R}^{2 \times 2}$ is given by:
\begin{equation}
    \boldsymbol{\Delta}=\left[\begin{array}{ll}
    \Delta_{t t} & \Delta_{t t^\prime} \\
    \Delta_{t^\prime t } & \Delta_{t^\prime t^\prime}
    \end{array}\right] .
\end{equation}

By introducing a Gaussian integral representation, the correlation $C_{tt'}$ can be expressed as:
\begin{equation}\label{ctt}
    \begin{aligned}
        C_{tt^\prime} =& \int Dz \int Dx  \phi(x\sqrt{\Delta_{tt} -\mid \Delta_{t t^\prime} \mid } +z \sqrt{\mid \Delta_{t t^\prime} \mid} ) \\
        &\int Dy \phi(y\sqrt{\Delta_{t^\prime t^\prime} -\mid \Delta_{t t^\prime} \mid } +z \sqrt{\mid \Delta_{t t^\prime} \mid} ) ,
    \end{aligned}
\end{equation}
where $Dz = \frac{\mathrm{d}z}{\sqrt{2\pi}} e^{-z^2/2}$. 

\blue{To further characterize the system's dynamics, we have introduced the autocorrelation function of the raw neural activity:
\begin{equation}
    \Delta_{tt^\prime} = \left\langle x(t)x(t^\prime) \right\rangle_\eta.
\end{equation}
By multiplying both sides of Eq.~\eqref{DMFT_Dynamics} and averaging over $\eta$, one obtains the following relation:
\begin{equation}
    (\partial_t +1 )(\partial_{t^\prime}+1)\Delta_{t t^\prime}= g^2 C_{t t^\prime}.
\end{equation}
Assuming that the system reaches a stationary regime, we consider the time-translationally invariant (TTI) limit where both $t,t^\prime \rightarrow \infty$, and only the time lag ($\tau=t-t'$) matters. In this regime, we can simplify the derivatives as $\partial_t=\partial_\tau , \quad \partial_{t^\prime}=-\partial_{\tau}$. In the TTI limit, we have the following motion equation:
\begin{equation}
    (1-\partial_\tau^2)\Delta_\tau = g^2 C_\tau.
\end{equation}
We further distinguish between the equal-time and time-lagged ($\tau$-lagged) correlations,  $\Delta_0$ and $\Delta_\tau=\Delta$, respectively.
In the steady-state limit, Equation~\eqref{ctt} leads to Eq.~(\ref{C_D_D0}) in the main text.}

Next, we define the effective potential:
    \begin{equation}
        V(\Delta;\Delta_0) = -\frac{1}{2}\Delta^2 +g^2 C_{\Phi}(\Delta;\Delta_0),
    \end{equation}
where the function $\Phi(x ) = \int_0^x \mathrm{d} y \phi(y) $ is introduced, and using the Price's theorem~\cite{Price-1958}: $\frac{\partial C_{\Phi}(\Delta;\Delta_0)}{\partial\Delta}=C_{\phi}$ (the subscript indicates
the non-linear function used to compute the correlation function, e.g., $\phi=\tanh$ in this paper), we get Eq.~(\ref{partial_tau^2_Delta}) in the main text.

The dynamics now depend on the initial condition $\Delta_0$. We next find a self-consistent solution of $\Delta_0$. According to the 
energy conservation, $V(\Delta_0;\Delta_0) = V(0;\Delta_0)$ ($\Delta=0$ in the second V due to the chaotic decorrelation), we derive:
\begin{equation}
    -\frac{\Delta_0^2}{2}+g^2\int Dz \Phi^2(\sqrt{\Delta_0}z) = g^2\left( \int Dz  \Phi(\sqrt{\Delta_0} z)\right)^2,
\end{equation}
from which the value of $\Delta_0$ can be obtained numerically. This leads to the root finding of $\mathcal{F}(\Delta_0;\Delta_0)=0$ in the main text.

Next, we provide details about deriving Eq.~\eqref{KE_TTI}. For the Fourier transform of $x(t)$, we define:
\begin{equation}
    \begin{aligned}
    &x(t) = \frac{1}{2\pi}\int \mathrm{d} \omega \hat{x}(\omega) e^{-i\omega t}, \\
    &\hat x(\omega) = \int \mathrm{d} t x(t)e^{i\omega t}.
    \end{aligned}
\end{equation}
Then the time derivative is given by
\begin{equation}
    \partial_t x(t) = \int  \frac{\mathrm{d} \omega}{2\pi} (-i \omega )\hat{x}(\omega)e^{-i \omega t}.
\end{equation}
The two-time correlation function is:
\begin{equation}
    \begin{aligned}
        \left\langle x(t)x(t^\prime)  \right\rangle 
        &=\left\langle  \int\int \frac{\mathrm{d} \omega}{2\pi} \frac{\mathrm{d} \omega^\prime}{2\pi} \hat{x}(\omega) \hat{x}^*(\omega^\prime) e^{-i \omega t }e^{i \omega^\prime t^\prime}\right\rangle   \\ 
        & =\int\int \frac{\mathrm{d} \omega}{2\pi} \frac{\mathrm{d} \omega^\prime}{2\pi} \left\langle   \hat{x}(\omega) \hat{x}^*(\omega^\prime)\right\rangle  e^{-i \omega t }e^{i \omega^\prime t^\prime},
    \end{aligned}
\end{equation}
where the correlation function in the frequency domain is given by
\begin{equation}\label{app-a}
    \begin{aligned}
        \left\langle\hat{x}(\omega) \hat{x}^*\left(\omega^{\prime}\right)\right\rangle 
        &=\left\langle\int d t x(t) e^{i \omega t} \int d t^{\prime} x\left(t^{\prime}\right) e^{-i \omega^{\prime} t^{\prime}}\right\rangle\\ 
        &=\int d t d t^{\prime} e^{i \omega t} e^{-i \omega^{\prime} t^{\prime}}\left\langle x(t) x\left(t^{\prime}\right)\right\rangle \\
        &=\int d t d t^{\prime} e^{i \omega t} e^{-i \omega^{\prime} t^{\prime}} \Delta\left(t,t^{\prime}\right)\\
        &\overset{\text{TTI}}{=} \int d t^{\prime} \int d \tau e^{i \omega\left(\tau+t^{\prime}\right)} e^{-i \omega^{\prime} t^{\prime}} \Delta(\tau) \\
        &=\int d \tau \Delta(\tau) e^{i \omega \tau} \int d t^{\prime} e^{i\left(\omega-\omega^{\prime}\right) t^{\prime}} \\
        &= 2\pi \delta(\omega-\omega^\prime ) \int \mathrm{d}\tau \Delta (\tau)e^{i \omega \tau} \\ 
        &= 2 \pi \delta(\omega-\omega^\prime)\hat{\Delta}(\omega).
    \end{aligned}
\end{equation}
To derive Eq.~\eqref{app-a},
we have used TTI, under which $\Delta_{tt'} \equiv \Delta(t - t')=\Delta(\tau)\equiv \Delta_\tau$ with $\tau = t - t'$, and thus the double integral over time can be rewritten using the change of variables $t = \tau + t'$. In addition, we used the identity
\begin{equation}
\delta(\omega - \omega') = \frac{1}{2\pi} \int \mathrm{d}t' e^{i(\omega - \omega') t'},
\end{equation}
which follows from the definition of the Dirac delta function in the frequency domain.
Moreover, we employed the fact that $x(t)$ is real-valued, which implies that its Fourier transform satisfies the Hermitian symmetry:
\begin{equation}
\hat{x}(-\omega) = \hat{x}^*(\omega),
\end{equation}
as can be seen from
\begin{equation}
    \hat{x}(-\omega) = \int x(t) e^{-i\omega t}\mathrm{d}t = \left( \int x(t) e^{i\omega t} \,\mathrm{d}t \right)^* = \hat{x}^*(\omega).
\end{equation}
Putting all these together, we arrive at Eq.~\eqref{KE_TTI}.

\section{Kinetic energy in the limit $g\to 1^{+}$}
\label{scaling_derivation}
\blue{Before performing the expansion, we notice that the derivatives of the potential $V(\Delta;\Delta_0)$ and the energy-conservation function $\mathcal{F}(\Delta;\Delta_0)$ are the same, since the two expressions differ only by a constant. This can be seen below.
\begin{equation}
    \begin{aligned}
        &\frac{\partial V(\Delta;\Delta_0)}{\partial \Delta} =\frac{\partial \mathcal{F}(\Delta;\Delta_0)}{\partial \Delta} =\\
        &\quad -\Delta + g^2 \int Dz\left[ \int Dx \phi(x\sqrt{\Delta_0-|\Delta|}+ z \sqrt{|\Delta|}) \right]^2,  \\
        \text{and}\quad
        &\frac{\partial^n V(\Delta;\Delta_0)}{\partial \Delta^n} =\frac{\partial^n \mathcal{F}(\Delta;\Delta_0)}{\partial \Delta^n } =-\frac{\partial^n}{\partial \Delta^n }\left(   \frac{\Delta^2}{2} \right) +\\
        &\quad g^2 \int Dz\left[ \int Dx \phi^{(n-1)}(x\sqrt{\Delta_0-|\Delta|}+ z \sqrt{|\Delta|}) \right]^2,
    \end{aligned}
\end{equation}
where the second equation is written for $n\geq1$, and $\phi^{(n)}=(d/dx)^n\phi(x)$.
We choose the activation function $\phi(x) = \tanh(x)$, which is an odd function. Consequently, all odd-order derivatives of $V(\Delta;\Delta_0)$ at $\Delta = 0$ vanish:
\begin{equation}\label{B2}
   \left. \frac{\partial^n V(\Delta;\Delta_0)}{\partial \Delta^n } \right|_{\Delta=0} = 0.
\end{equation}
To have Eq.~\eqref{B2}, we repeatedly used the Price's theorem~\cite{Price-1958}.}

\blue{Expanding $V(\Delta;\Delta_0)$ and $\mathcal{F}(\Delta;\Delta_0)$ for small $\Delta$, i.e., $|\Delta| \ll 1$, we obtain:
\begin{equation}
    \begin{aligned}
    \mathcal{F}(\Delta;\Delta_0)
    \thicksim
    & \mathcal{F}(0;\Delta_0)+\mathcal{F}^\prime (0;\Delta_0)\Delta +  \mathcal{F}^{\prime\prime} (0;\Delta_0)\frac{\Delta^2}{2} + \\ 
    &\mathcal{F}^{\prime\prime\prime} (0;\Delta_0)\frac{\Delta^3}{3!} +  \mathcal{F}^{\prime\prime\prime\prime}(0;\Delta_0)\frac{\Delta^4}{4!}+\mathcal{O}(\Delta^5) \\
    =&  \left( -1+g^2 \left[ \int Dx \phi^\prime(x\sqrt{\Delta_0}) \right]^2  \right)\frac{\Delta^2}{2}+ \\
    &g^2 \left[ \int Dx \phi^{\prime\prime\prime}(x\sqrt{\Delta_0}) \right]^2 \frac{\Delta^4}{4!}+\mathcal{O}(\Delta^5),
    \end{aligned}
\end{equation}
where $\mathcal{F}(0;\Delta_0)=0$,
and similarly,
\begin{equation}
    \begin{aligned}
    & V(\Delta;\Delta_0)
    \thicksim V(0;\Delta_0)+V^\prime (0;\Delta_0)\Delta + V^{\prime\prime} (0;\Delta_0)\frac{\Delta^2}{2} +\\ 
    &\quad\quad V^{\prime\prime\prime} (0;\Delta_0)\frac{\Delta^3}{3!} + V^{\prime\prime\prime\prime}(0;\Delta_0)\frac{\Delta^4}{4!}+\mathcal{O}(\Delta^5) \\
    &\quad =g^2  \left[ \int D x \Phi(x\sqrt{\Delta_0})\right]^2\\
    &+ \left( -1+g^2 \left[ \int Dx \phi^\prime(x\sqrt{\Delta_0}) \right]^2  \right)\frac{\Delta^2}{2} \\
    &\quad\quad +  g^2 \left[ \int Dx \phi^{\prime\prime\prime}(x\sqrt{\Delta_0}) \right]^2 \frac{\Delta^4}{4!}+\mathcal{O}(\Delta^5).
    \end{aligned}
\end{equation}
and finally from Eq.~\eqref{dV}, we have 
\begin{equation}
    \begin{aligned}
        \Gamma (\Delta,\Delta_0) &\thicksim
         \Gamma (0,\Delta_0)+\Gamma' (0,\Delta_0)\Delta +\Gamma'' (0,\Delta_0)\frac{\Delta^2}{2!}+\Gamma''' (0,\Delta_0)\frac{\Delta^3}{3!}\\
         &\quad+\Gamma'''' (0,\Delta_0)\frac{\Delta^4}{4!}+\mathcal{O}(\Delta^5)\\
         & =\left( -1+g^2 \left[ \int Dx \phi^\prime(x\sqrt{\Delta_0}) \right]^2  \right)\Delta + g^2 \left[ \int Dx \phi^{\prime\prime\prime}(x\sqrt{\Delta_0}) \right]^2 \frac{\Delta^3}{3!}+\mathcal{O}(\Delta^5).
    \end{aligned}
\end{equation}
}

\blue{For $\phi(x)=\tanh x$, we have the following Taylor series:
\begin{equation}
    \begin{aligned}
     &\Phi(x)=\ln\cosh(x)\thicksim \frac{x^2}{2}-\frac{x^4}{12} +\frac{x^6}{45}+\mathcal{O}(x^{8}), \\
     &\tanh(x) \thicksim x-\frac{x^3}{3}+\frac{2x^5}{15}-\frac{17}{315}x^7+\mathcal{O}(x^9),\\
    & \tanh'(x)\thicksim  1-x^2 +\frac{2x^4}{3}-\frac{17 x^6}{45} +\mathcal{O}(x^8),\\
     &\tanh''(x)\thicksim -2x +\frac{8x^3}{3} -\frac{34x^5}{15} +\mathcal{O}(x^7), \\
    & \tanh'''(x)\thicksim  -2 + 8x^2 -\frac{34x^4}{3}+\mathcal{O}(x^6).
    \end{aligned}
\end{equation}
Using Gaussian moments $\int Dz\,z^2=1$, $\int Dz\,z^4=3$, $\int Dz\,z^6=15$, $\int D z\,z^8 =105$ , we obtain
\begin{equation}
\begin{aligned}
    \left[ \int D x \Phi(x\sqrt{\Delta_0})\right]^2  
    & \thicksim
    \left[ \int D x\left(   \frac{x^2}{2}\Delta_0-\frac{x^4}{12}\Delta_0^2+\frac{x^6}{45}\Delta_0^3  \right)\right]^2 \\
    &\thicksim\frac{1}{4}\Delta_0^2 -\frac{1}{4}\Delta_0^3 + \frac{19}{48}\Delta_0^4-\frac{1}{6}\Delta_0^5, \\
    \left[ \int Dx \phi^\prime(x \sqrt{\Delta_0}) \right]^2 
    &\thicksim\left[ \int Dx \left( 1-x^2\Delta_0 +\frac{2}{3}x^4\Delta_0^2-\frac{17 x^6}{45}\Delta_0^3 \right)  \right]^2 \\
    &\thicksim 1-2\Delta_0 +5\Delta_0^2 - \frac{46}{3}\Delta_0^3,\\
    \left[ \int Dx \phi^{\prime\prime\prime}(x\sqrt{\Delta_0}) \right]^2 
    &\thicksim\left[ \int Dx \left(-2+8x^2{\Delta_0}-\frac{34}{3}x^4\Delta_0^2 \right) \right]^2\\
    &\thicksim 4- 32 \Delta_0.
\end{aligned}
\end{equation}
}

\blue{Thus, the expansions of $\mathcal{F}(\Delta; \Delta_0)$ ,$V(\Delta;\Delta_0)$ and $\Gamma(\Delta ;\Delta_0)$ read respectively:
\begin{equation}
    \begin{aligned}
    & \mathcal{F}(\Delta;\Delta_0)  \thicksim  \left[ -1+g^2\left(1-2\Delta_0 +5\Delta_0^2 -\frac{46}{3}\Delta_0^3\right) \right]\frac{\Delta^2}{2}+ \frac{g^2}{6}\Delta^4 -\frac{4}{3}g^2\Delta_0 \Delta^4, \\
    & V(\Delta;\Delta_0) \thicksim g^2 \left( \frac{1}{4}\Delta_0^2 -\frac{1}{4}\Delta_0^3 + \frac{19}{48}\Delta_0^4-\frac{1}{6}\Delta_0^5  \right)\\
    &\qquad\qquad\quad +  \left[ -1+g^2\left(1-2\Delta_0 +5\Delta_0^2-\frac{46}{3}\Delta_0^3\right) \right]\frac{\Delta^2}{2}+\frac{g^2 }{6} \Delta^4-\frac{4}{3}g^2\Delta_0 \Delta^4,\\
    & \Gamma(\Delta;\Delta_0) \thicksim \left[ -1+g^2\left(1-2\Delta_0 +5\Delta_0^2-\frac{46}{3}\Delta_0^3 \right) \right]\Delta +  \frac{g^2}{3} (2-16\Delta_0)\Delta^3.
    \end{aligned}
\end{equation}
These expressions are derived independently, but they are self-consistent, e.g., it can be verified that $\Gamma(\Delta;\Delta_0)=\partial_\Delta V(\Delta;\Delta_0)$ as expected.
}

\blue{From the energy conservation condition $\mathcal{F}(\Delta_0; \Delta_0) = V(\Delta_0;\Delta_0) - V(0;\Delta_0) = 0$, we obtain:
\begin{equation}\label{EC01}
    \left[ -1+g^2(1-2\Delta_0 +5\Delta_0^2 - \frac{46}{3}\Delta_0^3) \right]\frac{\Delta_0^2}{2}+g^2\frac{\Delta_0^4}{6} -g^2\frac{4}{3}\Delta_0^5 =0.
\end{equation}
We first set $g=1+\sigma$ where $\sigma \to 0^+$ and employ a perturbative expansion:
\begin{equation}
    \Delta_0 = A\sigma +B\sigma^2+C \sigma^3 + \mathcal{O}(\sigma^4).
\end{equation}
Equation~\eqref{EC01} is re-organized into the following form:
\begin{equation}
\frac{\Delta_0^2}{2} (g^2 -1 -2g^2\Delta_0 + \frac{16}{3}g^2 \Delta_0^2 -18g^2\Delta_0^3)  =  0,
\end{equation}
which implies that
\begin{equation}\label{EC02}
g^2 -1 - 2g^2 \Delta_0 + \frac{16}{3}g^2 \Delta_0^2 -18g^2\Delta_0^3 = 0,
\end{equation}
because $\Delta_0\neq0$.}

\blue{Now we can insert $g^2$ and $\Delta_0$ into Eq.~\eqref{EC02}, and we keep terms up to $\sigma^3$, resulting in the following form:
\begin{equation}
  \begin{aligned}
   &(\sigma^2+2 \sigma )+(1+2\sigma+\sigma^2 )
    \left[-2(A\sigma +B\sigma^2+C \sigma^3) +\frac{16}{3}(A^2\sigma^2 +2AB\sigma^3)-18 A^3\sigma^3 \right] = 0,\\
    &\text{reorganized into}\\
    &(2-2A)\sigma + (1-4A-2B+\frac{16}{3}A^2)\sigma^2 +(-2A-4B-2C+\frac{32}{3}A^2+\frac{32}{3}AB - 18A^3)\sigma^3 =0.
  \end{aligned}
\end{equation}
}

\blue{Because $\sigma\neq0$, we find the following equalities:
\begin{equation}
    \begin{aligned}
        &2-2A = 0  \quad\Rightarrow\quad A=1 \\
        & 1 -2B -4A +\frac{16A^2}{3} = 0\quad\Rightarrow\quad B=\frac{7}{6}\\ 
        &-2A-4B-2C+\frac{32}{3}A^2+\frac{32}{3}AB - 18A^3 = 0 \quad \Rightarrow\quad C=-\frac{7}{9}
    \end{aligned}
\end{equation}
Therefore, we conclude that
\begin{equation}\label{delta0}
    \Delta_0 \thicksim \sigma + \frac{7}{6}\sigma^2 -\frac{7}{9}\sigma^3 +\mathcal{O}(\sigma^4).
\end{equation}
}

\blue{Substituting the above $\Delta_0$ into the expansion of $\Gamma(\Delta;\Delta_0)$, we get:
\begin{equation}
    \begin{aligned}
     \Gamma(\Delta;\sigma) &\thicksim  
     \left[\sigma^2 +2\sigma -2(\sigma + \frac{19}{6}\sigma^2 +\frac{23}{9}\sigma^3)+5(\sigma^2 +\frac{13}{3}\sigma^3)-\frac{46}{3}\sigma^3\right] \Delta 
     + (\frac{2}{3} - 4\sigma )\Delta^3 \\
     & = (-\frac{1}{3}\sigma^2 +\frac{11}{9}\sigma^3)\Delta + (\frac{2}{3} - 4\sigma )\Delta^3.
    \end{aligned}
\end{equation}
}
\blue{Hence, the average kinetic energy is given by
\begin{equation}\label{Resgamma}
    \begin{aligned}
    \Gamma_0 &=\Gamma(\Delta_0(\sigma);\sigma)  
    \thicksim  (-\frac{1}{3}\sigma^2 +\frac{11}{9}\sigma^3)\Delta_0 + (\frac{2}{3} - 4\sigma )\Delta_0^3\\
    & \thicksim (-\frac{1}{3}\sigma^2 +\frac{11}{9}\sigma^3)(\sigma+\frac{7}{6}\sigma^2) + (\frac{2}{3} - 4\sigma )(\sigma^3 +\frac{21}{6}\sigma^4)\\
    & \thicksim  \frac{1}{3}\sigma^3 -\frac{5}{6}\sigma^4+O(\sigma^5).
    \end{aligned}
\end{equation}
}

\blue{The same result can also be obtained by directly expanding $\Gamma(\Delta_0)$:
\begin{equation}
    \begin{aligned}
    \Gamma_0 &\equiv \Gamma(\Delta_0;\Delta_0) = g^2C(\Delta_0;\Delta_0) - \Delta_0 = g^2 \int Dz \left[  \phi (z\sqrt{|\Delta_0|}) \right]^2 - \Delta_0 ,
    \label{Gamma_0_app}
    \end{aligned}
\end{equation}
Expansion of $\Gamma_0$ for small $\Delta_0$, and substituting $\Delta_0 \thicksim \sigma + \frac{7}{6}\sigma^2 -\frac{7}{9}\sigma^3$, we get
\begin{equation}
    \begin{aligned}
        \Gamma_0 &\thicksim g^2 \left[ \int Dz \left(z \sqrt{\Delta_0 }- \frac{z^3}{3}\sqrt{\Delta_0^3} +\frac{2z^5}{15}\sqrt{\Delta_0^5} \right)^2  \right] - \Delta_0 \\
        & \thicksim g^2 \left[ \int Dz \left( z^2\Delta_0 -\frac{2}{3}z^4 \Delta_0^2 +\frac{17}{45}z^6\Delta_0^3-\frac{62}{315}z^8 \Delta_0^4  \right)  \right] - \Delta_0 \\
        & = (g^2-1)\Delta_0 -2g^2\Delta_0^2 +\frac{17}{3}g^2\Delta_0^3 - \frac{62}{3}g^2\Delta_0^4 \\
                & \thicksim 
        (2\sigma^2 +\frac{10}{3}\sigma^3-\frac{7}{18}\sigma^4 )+ (-2\sigma^2 -3\sigma^3 -\frac{4}{9}\sigma^4)+ \mathcal{O}(\sigma^5)\\
        &= \frac{1}{3}\sigma^3 -\frac{5}{6}\sigma^4+\mathcal{O}(\sigma^5).
    \end{aligned}
\end{equation}
This result is consistent with Eq.~\eqref{Resgamma}.
}

\blue{We finally remark that in Ref.~\cite{PRE-2018}, a different form of dynamics equation in comparison to Eq.~\eqref{neuron_dynamics} was used. The dynamics read
  \begin{equation}\label{Dyn2}
    \dot h_i(t)=-h_i(t)+\sum_{j=1}^N J_{ij}\phi(g h_j(t)),
    \qquad J_{ij}\sim \mathcal N\!\left(0,\frac{1}{N}\right).
    \end{equation}
Both dynamics share the same chaos transition point ($g_c=1$). However, the critical behavior just above $g_c$ is different when higher-order corrections are taken. We just show the main results of key quantities here.
}

\blue{First, the corresponding $\mathcal{F}$ function [actually $V(\Delta;\Delta_0)$ in Ref.~\cite{PRE-2018}, see Eq. (50) there] reads
\begin{equation}
    \mathcal{F}({\Delta;\Delta_0}) = -\frac{1}{2}\Delta^2 + \frac{1}{g^2}\int D z \left[\int D x \Phi \left(gx \sqrt{\Delta_0 - |\Delta|} + g z \sqrt{\left\vert \Delta \right\vert }\right)\right]^2 - \frac{1}{g^2}\left[\int D x \Phi (gx \sqrt{\Delta_0})\right]^2,
\end{equation}
and the function $\Gamma(\Delta;\Delta_0)$ is given as
\begin{equation}
    \begin{aligned}
    \Gamma(\Delta;\Delta_0) &= \frac{\partial V(\Delta;\Delta_0)}{\partial \Delta }=C(\Delta;\Delta_0) - \Delta   
    = -\Delta +  \int D z \left[ \int D x \phi (gx \sqrt{\Delta_0 -\left\vert \Delta \right\vert } +gz \sqrt{\left\vert \Delta \right\vert })\right]^2.
    \end{aligned}
\end{equation}
}

\blue{By applying the Price's theorem (see Eq. (C8) in Ref.~\cite{PRE-2018}), and considering a small $\Delta$ expansion, we get
 \begin{equation}
    \begin{aligned}
      \mathcal{F}(\Delta;\Delta_0)
      &\sim
      \left[
      -1+g^2-2g^4\Delta_0+5g^6\Delta_0^2-\frac{46}{3}g^8\Delta_0^3
      \right]\frac{\Delta^2}{2}
      +\frac{g^6}{6}\Delta^4
      -\frac{4}{3}g^8\Delta_0\Delta^4.
      \label{eq:F_expansion}
      \end{aligned}
  \end{equation}
  Similarly, we get
  \begin{equation}
    \begin{aligned}
      \Gamma(\Delta;\Delta_0)
      &\sim
      \left[
      -1+g^2-2g^4\Delta_0+5g^6\Delta_0^2-\frac{46}{3}g^8\Delta_0^3
      \right]\Delta
      +
      \left(
      \frac{2}{3}g^6-\frac{16}{3}g^8\Delta_0
      \right)\Delta^3.
      \end{aligned}
 \end{equation}
 }

\blue{From the energy conservation condition $\mathcal{F}(\Delta_0;\Delta_0) =0$, and further assuming $\Delta_0=A\sigma+B\sigma^2+C\sigma^3$, we get 
  \begin{equation}
  \Delta_0\sim \sigma-\frac{5}{6}\sigma^2-\frac{1}{9}\sigma^3+O(\sigma^4).
  \end{equation}
  Therefore, the term $O(\sigma^2)$ for the dynamics considered in Ref.~\cite{PRE-2018} is already different from ours [see Eq.~\eqref{delta0}]. We remark that the second-order term given in Ref.~\cite{PRE-2018} is incorrect (probably a typo). Inserting the correct form into $\mathcal{F}$ function, we get
\begin{equation}
  \begin{aligned}
   & \mathcal{F}(\Delta;\Delta_0)  \sim (-1+g^2-2g^4 \Delta_0 + 5g^6 \Delta_0^2 )\frac{\Delta^2}{2}+g^6\frac{\Delta^4}{6} \\
   & \sim (2\sigma +\sigma^2-2(1+4\sigma)(\sigma-\frac{5}{6}\sigma^2)+5\sigma^2)\frac{\Delta^2}{2} + \frac{1}{6}\Delta^4 \\
   & = -\frac{\sigma^2}{6}\Delta^2+\frac16\Delta^4,
  \end{aligned}
  \end{equation}
which is consistent with Eq. (124) in Ref.~\cite{PRE-2018}.
}

\blue{Setting $\Delta=\Delta_0$ in $\Gamma(\Delta;\Delta_0)$ and inserting the perturbative expression of $\Delta_0$, we get the kinetic energy for the dynamics~\eqref{Dyn2}:
 \begin{equation}
  \Gamma_0=\Gamma(\Delta_0)
  \sim
  \left(
  -\frac{1}{3}\sigma^2+\frac{11}{9}\sigma^3
  \right)\Delta_0
  +
  \left(
  \frac{2}{3}-\frac{4}{3}\sigma
  \right)\Delta_0^3.
  \end{equation}
  Using
  \begin{equation}
  \Delta_0\sim \sigma-\frac{5}{6}\sigma^2-\frac{1}{9}\sigma^3,
  \end{equation}
  we obtain
  \begin{align}
  \Gamma_0
  &\sim
  \left(
  -\frac{1}{3}\sigma^2+\frac{11}{9}\sigma^3
  \right)\left(
  \sigma-\frac{5}{6}\sigma^2
  \right)
  +
  \left(
  \frac{2}{3}-\frac{4}{3}\sigma
  \right)\left(
  \sigma^3-\frac{5}{2}\sigma^4
  \right) \nonumber\\
  &\sim
  \frac{1}{3}\sigma^3-\frac{3}{2}\sigma^4+O(\sigma^5).
  \end{align}
  Thus,
  \begin{equation}
  \Gamma_0\sim \frac{1}{3}\sigma^3-\frac{3}{2}\sigma^4+O(\sigma^5).
  \end{equation}
  }
  
\begin{figure}[t]
    \centering
    \includegraphics[width=\textwidth]{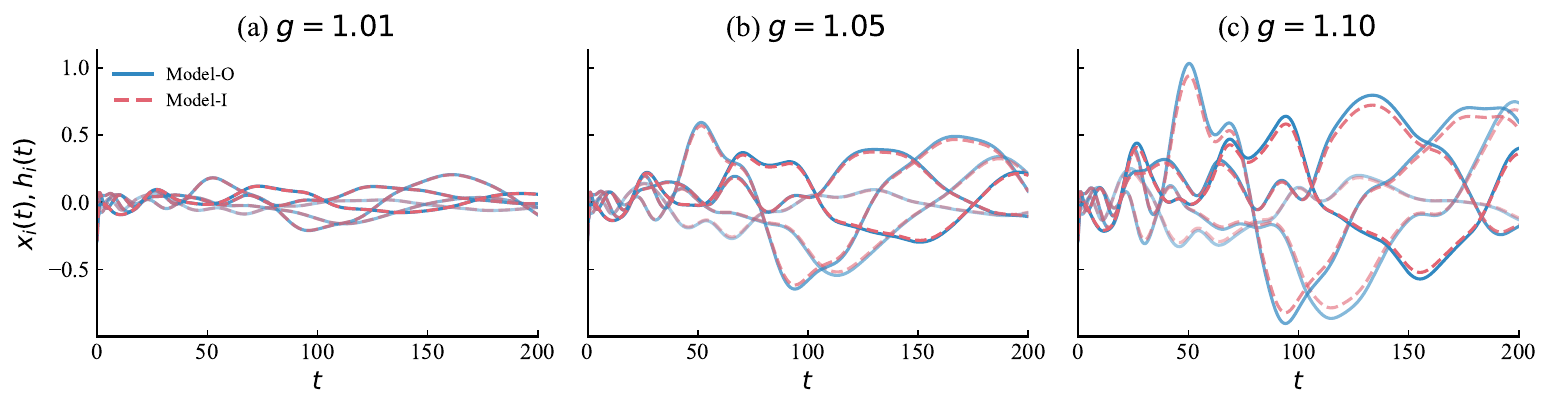}
    \caption{
    \blue{Comparison of dynamics trajectories for Model-O (this paper) and Model-I (Ref.~\cite{PRE-2018}) using the same random coupling matrix and the same initial condition. 
    The network size is \(N=5000\), the total simulation time is \(200\) steps, and the unit of time is \(dt=0.1\).
    In each panel, the trajectories of five neurons are shown for both models with different colors indicating the two dynamics.
    Panels (a)--(c) correspond to \(g=1.01\), \(g=1.05\), and \(g=1.10\), respectively.
    The plot shows that the two models have the same critical point \(g_c=1\), but their detailed trajectories deviate from each other in the chaos regime especially away from the chaos onset.}
    \label{comp}
    }
\end{figure}

  \blue{We conclude that both dynamics yield the same leading order, but different forms of higher-order correction, as proved in numerical simulations (Fig.~\ref{comp}).}

\section{Algorithmic implementation of the DMFT equation}
\label{Algorithm}
We add here the details about the implementation of the single-degree-of-freedom DMFT equations, as shown in Algorithm~\ref{alg:dmft}. For more implementation details of the dynamical mean-field theory, readers may refer to a pedagogical introduction~\cite{Zou-2024}. More efficient methods are presented in previous works~\cite{PRE-2018,PRX-2021}.

    \begin{algorithm}[H]
        \caption{DMFT Dynamics Simulation}
        \label{alg:dmft}
        \begin{algorithmic}[1]
        \State Draw $M$ samples of noise trajectories $\eta_t^{m,(0)}$ from a multivariate Gaussian distribution with zero mean and covariance $\langle \eta(t)\eta(t') \rangle$.
        \State Initialize $x^m(0) \sim \mathcal{N}(0, 1)$ for $m = 1, \dots, M$; set total time length $S = T/dt$ ($dt$ is a small increment).
        \For{$c = 1$ to $N_c$} \Comment{DMFT iteration}
            \For{$t = 0$ to $S-1$}
                \State $x^m(t{+}1) \gets x^m(t) + \mathrm{d}t \cdot \left[-x^m(t) + \eta_t^{m,(c{-}1)}\right]$
            \EndFor
            \State Compute $\phi^m(t) \gets \tanh(x^m(t))$
            \State Estimate correlation: $C_{tt'} \gets \frac{1}{M} \sum\limits_{m=1}^M \phi^m(t) \phi^m(t')$
            \State Estimate variance: $\Delta_{tt'} \gets \frac{1}{M} \sum\limits_{m=1}^M x^m(t) x^m(t')$
            \State Draw $\eta_t^{m,(c)} \sim \mathcal{N}(0, g^2 C_{tt'})$ \Comment{Update noise with current covariance}
        \EndFor
        \State \textbf{Output:} Final $\Delta_{tt'}$ and $C_{tt'}$
        \end{algorithmic}
        \end{algorithm}

%\bibliography{ref}

\begin{thebibliography}{10}

\bibitem{Chaos-1988}
H.~Sompolinsky, A.~Crisanti, and H.~J. Sommers.
\newblock Chaos in random neural networks.
\newblock {\em Phys. Rev. Lett.}, 61:259--262, 1988.

\bibitem{PRE-2018}
A.~Crisanti and H.~Sompolinsky.
\newblock Path integral approach to random neural networks.
\newblock {\em Phys. Rev. E}, 98:062120, 2018.

\bibitem{Roy-2019}
F~Roy, G~Biroli, G~Bunin, and C~Cammarota.
\newblock Numerical implementation of dynamical mean field theory for
  disordered systems: application to the lotka–volterra model of ecosystems.
\newblock {\em Journal of Physics A: Mathematical and Theoretical},
  52(48):484001, 2019.

\bibitem{CMP-2024}
Leticia~F. Cugliandolo.
\newblock Recent applications of dynamical mean-field methods.
\newblock {\em Annual Review of Condensed Matter Physics}, 15(Volume 15,
  2024):177--213, 2024.

\bibitem{Zou-2024}
Wenxuan Zou and Haiping Huang.
\newblock {Introduction to dynamical mean-field theory of randomly connected
  neural networks with bidirectionally correlated couplings}.
\newblock {\em SciPost Phys. Lect. Notes}, page~79, 2024.

\bibitem{Carles-2024}
Carles Martorell, Rub{\'e}n Calvo, Alessia Annibale, and Miguel~A Munoz.
\newblock Dynamically selected steady states and criticality in non-reciprocal
  networks.
\newblock {\em Chaos, Solitons \& Fractals}, 182:114809, 2024.

\bibitem{Exact-2025}
Francesco Ferraro, Christian Grilletta, Amos Maritan, Samir Suweis, and Sandro
  Azaele.
\newblock Exact solution of dynamical mean-field theory for a linear system
  with annealed disorder.
\newblock {\em Journal of Statistical Mechanics: Theory and Experiment},
  2025(2):023301, 2025.

\bibitem{Qiu-2025}
Junbin Qiu and Haiping Huang.
\newblock An optimization-based equilibrium measure describing fixed points of
  non-equilibrium dynamics: application to the edge of chaos.
\newblock {\em Commun. Theor. Phys.}, 77(3):035601, 2025.

\bibitem{EoC-1990}
Chris~G. Langton.
\newblock Computation at the edge of chaos: Phase transitions and emergent
  computation.
\newblock {\em Physica D: Nonlinear Phenomena}, 42(1):12--37, 1990.

\bibitem{EoC-2004}
Nils Bertschinger and Thomas Natschl{\"a}ger.
\newblock Real-time computation at the edge of chaos in recurrent neural
  networks.
\newblock {\em Neural computation}, 16(7):1413--1436, 2004.

\bibitem{Maass-2009}
Dean~V. Buonomano and Wolfgang Maass.
\newblock State-dependent computations: spatiotemporal processing in cortical
  networks.
\newblock {\em Nature Reviews Neuroscience}, 10(2):113--125, 2009.

\bibitem{Takasu-2025}
Shotaro Takasu and Toshio Aoyagi.
\newblock Neuronal correlations shape the scaling behavior of memory capacity
  and nonlinear computational capability of reservoir recurrent neural
  networks.
\newblock {\em Physical Review Research}, 7(4):043083, 2025.

\bibitem{Abbott-2009}
David Sussillo and Larry~F Abbott.
\newblock Generating coherent patterns of activity from chaotic neural
  networks.
\newblock {\em Neuron}, 63(4):544--557, 2009.

\bibitem{Kadmon-2015}
Jonathan Kadmon and Haim Sompolinsky.
\newblock Transition to chaos in random neuronal networks.
\newblock {\em Phys. Rev. X}, 5:041030, 2015.

\bibitem{Mastrogiuseppe-2018}
Francesca Mastrogiuseppe and Srdjan Ostojic.
\newblock Linking connectivity, dynamics, and computations in low-rank
  recurrent neural networks.
\newblock {\em Neuron}, 99(3):609--623, 2018.

\bibitem{Vyas-2020}
Saurabh Vyas, Matthew~D. Golub, David Sussillo, and Krishna~V. Shenoy.
\newblock Computation through neural population dynamics.
\newblock {\em Annual Review of Neuroscience}, 43:249--275, 2020.

\bibitem{Huang-2024}
Haiping Huang.
\newblock Eight challenges in developing theory of intelligence.
\newblock {\em Front. Comput. Neurosci}, 18:1388166, 2024.

\bibitem{Ostojic-2024}
Srdjan Ostojic and Stefano Fusi.
\newblock Computational role of structure in neural activity and connectivity.
\newblock {\em Trends in Cognitive Sciences}, 28(7):677--690, 2024.

\bibitem{Helias-2025}
Jakob Stubenrauch, Christian Keup, Anno~C. Kurth, Moritz Helias, and Alexander
  van Meegen.
\newblock Fixed point geometry in chaotic neural networks.
\newblock {\em Phys. Rev. Res.}, 7:023203, 2025.

\bibitem{HWZ-2025}
Weizhong Huang and Haiping Huang.
\newblock Freezing chaos without synaptic plasticity.
\newblock {\em Phys. Rev. E}, 112:044227, Oct 2025.

\bibitem{Wang-2024}
Shishe Wang and Haiping Huang.
\newblock How high dimensional neural dynamics are confined in phase space.
\newblock {\em arXiv:2410.19348}, 2024.

\bibitem{Yang-2025}
Xiaoyu Yang, Giancarlo~La Camera, and Gianluigi Mongillo.
\newblock On the relationship between equilibria and dynamics in large, random
  neuronal networks.
\newblock {\em arXiv:2510.19091}, 2025.

\bibitem{Urbani-2025}
Samantha~J. Fournier, Alessandro Pacco, Valentina Ros, and Pierfrancesco
  Urbani.
\newblock Nonreciprocal interactions and high-dimensional chaos: Comparing
  dynamics and statistics of equilibria in a solvable class of models.
\newblock {\em Phys. Rev. E}, 113:044139, 2026.

\bibitem{NN-2013}
David Sussillo and Omri Barak.
\newblock Opening the black box: Low-dimensional dynamics in high-dimensional
  recurrent neural networks.
\newblock {\em Neural Computation}, 25:626--649, 2013.

\bibitem{Yu-2025}
Zhendong Yu, Weizhong Huang, and Haiping Huang.
\newblock Neural langevin machine: a local asymmetric learning rule can be
  creative.
\newblock {\em arXiv:2506.23546}, 2025.

\bibitem{Clark-2023}
David~G. Clark, L.~F. Abbott, and Ashok Litwin-Kumar.
\newblock Dimension of activity in random neural networks.
\newblock {\em Phys. Rev. Lett.}, 131:118401, 2023.

\bibitem{PRR-2023}
Rainer Engelken, Fred Wolf, and L.~F. Abbott.
\newblock Lyapunov spectra of chaotic recurrent neural networks.
\newblock {\em Phys. Rev. Res.}, 5:043044, 2023.

\bibitem{Helias-2020}
M.~Helias and D.~Dahmen.
\newblock {\em Statistical field theory for neural networks}.
\newblock Springer, Berlin, 2020.

\bibitem{Huang-2022}
Haiping Huang.
\newblock {\em Statistical Mechanics of Neural Networks}.
\newblock Springer, Singapore, 2022.

\bibitem{Price-1958}
R~Price.
\newblock A useful theorem for nonlinear devices having gaussian inputs.
\newblock {\em IRE Transactions on Information Theory}, 4(2):69--72, 1958.

\bibitem{Risken-1996}
Hannes Risken.
\newblock {\em The Fokker-Planck Equation: Methods of Solution and
  Applications}.
\newblock Springer-Verlag Berlin, Berlin, 1996.

\bibitem{Van-2007}
N.G.~Van Kampen.
\newblock {\em Stochastic Processes in Physics and Chemistry}.
\newblock 3rd ed., North-Holland Personal Library, North-Holland, Amsterdam,
  2007.

\bibitem{Fiete-2019}
Rishidev Chaudhuri, Berk Ger{\c c}ek, Biraj Pandey, Adrien Peyrache, and Ila
  Fiete.
\newblock The intrinsic attractor manifold and population dynamics of a
  canonical cognitive circuit across waking and sleep.
\newblock {\em Nature Neuroscience}, 22(9):1512--1520, 2019.

\bibitem{Fusi-2023}
Ramon Nogueira, Chris~C. Rodgers, Randy~M. Bruno, and Stefano Fusi.
\newblock The geometry of cortical representations of touch in rodents.
\newblock {\em Nature Neuroscience}, 26(2):239--250, 2023.

\bibitem{Miller-2025}
Mauro~M. Monsalve-Mercado, Gabriel~M. Stine, Michael~N. Shadlen, and Kenneth~D.
  Miller.
\newblock The geometry of the neural state space of decisions.
\newblock {\em bioRxiv}, 2025.

\bibitem{Clark-2024}
David~G. Clark and L.~F. Abbott.
\newblock Theory of coupled neuronal-synaptic dynamics.
\newblock {\em Phys. Rev. X}, 14:021001, 2024.

\bibitem{Du-2024}
Wenkang Du and Haiping Huang.
\newblock Synaptic plasticity alters the nature of the chaos transition in
  neural networks.
\newblock {\em Phys. Rev. E}, 112:054208, 2025.

\bibitem{Jstat-2023}
Samantha~J Fournier and Pierfrancesco Urbani.
\newblock Statistical physics of learning in high-dimensional chaotic systems.
\newblock {\em Journal of Statistical Mechanics: Theory and Experiment},
  2023(11):113301, 2023.

\bibitem{ZLR-2025}
Liru Zhang.
\newblock https://github.com/Zliru/RNN-Kinetic, 2025.

\bibitem{PRX-2021}
Christian Keup, Tobias K\"uhn, David Dahmen, and Moritz Helias.
\newblock Transient chaotic dimensionality expansion by recurrent networks.
\newblock {\em Phys. Rev. X}, 11:021064, 2021.

\end{thebibliography}

\end{document}